\documentclass[floatfix,showpacs,preprintnumbers,nofootinbib,amsmath,amssymb]{revtex4}
\usepackage{graphicx}
\usepackage{dcolumn}
\usepackage{bm}

\newcommand{\Real}{\mathbb{R}}

\newcommand{\grad}{{\bf grad}\,}      
\newcommand{\curl}{{\bf curl}\,}      
\newcommand{\divr}{{\bf div}\,}       

\newtheorem{lemma}{Lemma}
\newtheorem{theorem}{Theorem}
\newcommand{\proof}{\noindent {\bf Proof. }}
\newcommand{\qed}{\hfill $\fbox{\hspace{0.3mm}}$ \vspace{.3cm}} 

\newcommand{\hateq}{\; \hat{=}\; }

\begin{document}
\title{Boundary conditions for the Baumgarte-Shapiro-Shibata-Nakamura 
formulation of Einstein's field equations}
\author{Dar\'\i o N\'u\~nez$^{1}$ and Olivier Sarbach$^{2}$}
\affiliation{$^{1}$Instituto de Ciencias Nucleares, Universidad
Nacional Aut\'onoma de M\'exico, Apartado postal 70-543, Ciudad Universitaria, 
04510 M\'exico, Distrito Federal, M\'exico.\\
$^{2}$Instituto de F\'\i sica y Matem\'aticas,
Universidad Michoacana de San Nicol\'as de Hidalgo\\
Edificio C-3, Ciudad Universitaria, 58040 Morelia,
Michoac\'an, M\'exico.} 
\email{nunez@nucleares.unam.mx, sarbach@ifm.umich.mx}

\begin{abstract}
We discuss the initial-boundary value problem for the Baumgarte-Shapiro-Shibata-Nakamura evolution system of Einstein's field equations which has been used extensively in numerical simulations of binary black holes and neutron stars. We specify nine boundary conditions for this system with the following properties: (i) they impose the momentum constraint at the boundary, which is shown to preserve all the constraints throughout evolution, (ii) they approximately control the incoming gravitational degrees of freedom by specifying the Weyl scalar $\Psi_0$ at the boundary, (iii) they control the gauge freedom by requiring a Neumann boundary condition for the lapse, by setting the normal component of the shift to zero, and by imposing a Sommerfeld-like condition on the tangential components of the shift, (iv) they are shown to yield a well-posed problem in the limit of weak gravity. Possible numerical applications of our results are also discussed briefly.
\end{abstract}

\date{\today}


\maketitle

\section{Introduction}

Many of the boundaries introduced in the modeling of physical problems are not part of the system. In the numerical Cauchy evolution of a gravitational system, for instance, one is faced with the finiteness of the numerical grid which usually leads to an evolution scheme on a spacetime domain with artificial initial and outer boundary surfaces. Initial conditions are chosen such that they represent as accurately as possible the physical system at one moment of time. The outer boundary surface in turn needs to be transparent to the system in the sense that the conditions imposed on this surface mimic as well as possible the evolution on an infinite, asymptotically flat domain. This leads to the construction of absorbing outer boundary conditions which can be defined by the requirement that they yield a well-posed initial-boundary value problem (IBVP) and minimize spurious reflections from the outer boundary. Furthermore, the boundary conditions should be chosen such that they preserve the constraints throughout evolution.

There has already been a large amount of work on outer boundary conditions in General Relativity, see \cite{oS07} for a recent review. Absorbing boundary conditions which preserve the constraints and lead to a well-posed IBVP have first been constructed for a tetrad formulation \cite{hFgN99} and more recently
\cite{hKjW06,hKoRoSjW07,mRoRoS07,hKoRoSjW09} also for the harmonic formulation of
Einstein's vacuum field equations. In particular, error estimates for spurious reflections and  higher-order absorbing boundary conditions have been given in \cite{lBoS06,lBoS07} and implemented and tested numerically in \cite{oRlLmS07,oRlBmShP09}.

However, these results are not yet applicable to the Baumgarte-Shapiro-Shibata-Nakamura (BSSN) formulation \cite{mStN95,tBsS99} of Einstein's field equations on which
many numerical codes for simulating binary black holes or neutron stars are based. So far, such codes use ad-hoc ``radiative boundary condition'' (see, for example, Ref. \cite{Alcubierre-Book}), which are imposed on all the evolution variables. Although these conditions seem very easy to implement numerically, they are unlikely to yield a well-posed Cauchy evolution or to preserve the constraints. This results in a lack of either accuracy or efficiency since the outer boundary has to be pushed very far away in the wave zone in such a way that the boundary surface is causally disconnected from the region where physics is extracted.

In this article we construct absorbing boundary conditions for the BSSN system which preserve the constraints and discuss the well-posedness of the resulting initial-boundary
value problem. Previous work on specifying boundary conditions for the BSSN system has been carried out in \cite{hBoS04}, where the BSSN system for a rather general class \cite{oSmT01} of gauge conditions on the lapse and a fixed shift tangent to the boundary is shown to be reducible to a first-order symmetric hyperbolic system (FOSH). Then, boundary conditions are given to the six incoming characteristic fields, resulting in a well-posed IBVP. However, the conditions presented in \cite{hBoS04} do not preserve the constraints. As a consequence, constraint-violating modes may be generated at the boundary surface. The work in \cite{cGjM04b} on the other hand, formulates constraint-preserving boundary conditions for the BSSN system but the well-posedness of the resulting IBVP in the absorbing case is not established. Furthermore, numerical simulations of binary black hole mergers require more general gauge conditions on the shift than the ones considered for the IBVPs formulated in \cite{hBoS04,cGjM04b}.

Here, we present a new set of boundary conditions for the BSSN system with the "hyperbolic $K$-driver" condition for the lapse and the ``hyperbolic Gamma-driver'' condition for the shift vector \cite{mAetal03} which is currently used in numerical simulations of binary black holes. The main properties of our boundary conditions are: (i)
they preserve the constraints throughout evolution, (ii) they control the Weyl scalar $\Psi_0$ at the boundary, a condition that yields small spurious reflections of gravitational radiation \cite{lBoS06,oRlLmS07}, (iii) they provide conditions on lapse and shift in the form of a Neumann condition for the lapse, a Dirichlet condition for the normal component of the shift and a Sommerfeld-like boundary condition on the tangential components of the shift, (iv) they are shown to yield a well-posed IBVP in the case of weak gravitational fields, where the equations may be linearized about Minkowski spacetime. If the outer boundary is placed in the weak field zone, this should be a reasonable assumption, and in view of the results in Ref. \cite{hBoS04} about the well-posedness of the Cauchy problem without boundaries one might except that our boundary conditions yield a well-posed IBVP in the nonlinear case as well.

Our boundary conditions are summarized in Sec.~\ref{Sec:MainResult} along with the BSSN evolution equations and constraints, and in the rest of the article the derivation of these boundary conditions is presented. The derivation is based on the analysis of small amplitude, high-frequency perturbations of the system which gives rise to a linear system on the half space $x > 0$ with planar boundary at $x=0$. Boundary conditions are first constructed for the linearized system and then extrapolated to the nonlinear case. The linear system is studied in Sec.~\ref{Sec:LinSystem} using the methods described in \cite{oRoS05,gNoS06} where boundary conditions are constructed in three steps. In a first step, the propagation of the {\em gauge-invariant quantities}, consisting of the constraint variables and the linearized Weyl curvature is analyzed. It is shown that this system yields a FOSH evolution system. We specify boundary conditions for this system which yield a well-posed IBVP, guarantee  the propagation of the constraint fields and also control the Weyl scalar $\Psi_0$. The second step consists in controlling the {\em gauge degrees of freedom}. The gauge functions, the lapse $\alpha$, and the shift vector $\beta^i$ are free to be chosen as better fits the problem. That freedom, however, is diminished by several practical requirements. The hyperbolic $K$-driver and Gamma-driver mentioned above are used in most BSSN formulations. We show that this system can be cast into FOSH form, and specify well-posed boundary conditions. In particular, the boundary conditions require that the normal component of the shift is zero, a condition that is important in order to ensure that the number of ingoing characteristic fields is constant throughout evolution, as explained in the next section. The final step consists in reconstructing the BSSN variables and to show that the initial data and our boundary conditions give rise to unique solutions of the IBVP. The residual gauge freedom in the linearized IBVP is also analyzed in Sec.~\ref{Sec:LinSystem}, and it is shown to be parametrized by two functions and three vector fields on the initial slice and one transverse vector field on the boundary.

Next, in Sec.~\ref{Sec:CP} we show that our boundary conditions do, in fact, preserve the constraints in the nonlinear case. This means that solutions of the nonlinear BSSN system satisfying our boundary conditions and the initial constraints satisfy the constraints everywhere on the spacetime domain. We present a brief discussion of a possible numerical implementation in Sec.~\ref{Sec:Numer} and draw conclusions in Sec.~\ref{Sec:Conc}. Since our methods are based on the theory of FOSH systems with maximal dissipative boundary conditions, we summarize the relevant results of this theory  in an appendix.

We finish our introduction by mentioning that there are alternative methods to cope with the problem of representing an infinite domain on a finite numerical grid. Examples are the Cauchy-characteristic matching method \cite{jW01} and the hyperboloidal initial value problem where a compactification along  hyperboloidal surfaces in a scri-fixing gauge allows to describe the gravitational waveform at null infinity in an unambiguous way, and no outer boundary conditions are needed. We refer the reader to Refs. \cite{Zeng08,vMoR09,oR09,lBhPjB09,Zeng09,aZmT09} for recent developments and the original references on the hyperboloidal initial value problem. While still under development, it constitutes a very promising approach for the computation of gravitational waves emitted by isolated systems.

\section{Boundary conditions for the BSSN system}
\label{Sec:MainResult}

The evolution equations we are referring to in our work are the BSSN evolution equations with the hyperbolic $K$-driver and Gamma-driver gauge conditions as described in \cite{mAetal03} with the exception of the advection terms. Using the notation in \cite{hBoS04} this system reads
\begin{eqnarray}
\hat{\partial}_0 \alpha &=& -\alpha^2 f(\alpha,\phi,x^\mu) (K - K_0(x^\mu)),
\label{Eq:BSSN1}\\
\hat{\partial}_0 K &=& -e^{-4\phi}\left[ 
 \tilde{D}^i\tilde{D}_i \alpha + 2\partial_i\phi \cdot\tilde{D}^i\alpha \right]
 + \alpha\left( \tilde{A}^{ij}\tilde{A}_{ij} + \frac{1}{3} K^2 \right)
 - \alpha S,
\label{Eq:BSSN2}\\
\hat{\partial}_0 \beta^i &=& \alpha^2 G(\alpha,\phi,x^\mu) B^i,
\label{Eq:BSSN3}\\
\hat{\partial}_0 B^i &=& e^{-4\phi} H(\alpha,\phi,x^\mu)
  \hat{\partial}_0\tilde{\Gamma}^i - \eta^i(B^j,\alpha,x^\mu)
\label{Eq:BSSN4}\\
\hat{\partial}_0 \phi &=& -\frac{\alpha}{6}\, K + \frac{1}{6}\partial_k\beta^k,
\label{Eq:BSSN5}\\
\hat{\partial}_0 \tilde{\gamma}_{ij} &=& -2\alpha\tilde{A}_{ij} 
 + 2\tilde{\gamma}_{k(i}\partial_{j)}\beta^k 
 - \frac{2}{3}\tilde{\gamma}_{ij}\partial_k\beta^k ,
\label{Eq:BSSN6}\\
\hat{\partial}_0 \tilde{A}_{ij} &=& e^{-4\phi}\left[ 
 \alpha\tilde{R}_{ij} + \alpha R^\phi_{ij} - \tilde{D}_i\tilde{D}_j\alpha 
  + 4\partial_{(i}\phi\cdot\tilde{D}_{j)}\alpha\right]^{TF}
\nonumber\\
 &+& \alpha K\tilde{A}_{ij} - 2\alpha\tilde{A}_{ik}\tilde{A}^k_{\; j}
  + 2\tilde{A}_{k(i}\partial_{j)}\beta^k 
  - \frac{2}{3}\tilde{A}_{ij}\partial_k\beta^k
  - \alpha e^{-4\phi} \hat{S}_{ij} ,
\label{Eq:BSSN7}\\
\hat{\partial}_0\tilde{\Gamma}^i &=& 
 \tilde{\gamma}^{kl}\partial_k\partial_l\beta^i
 + \frac{1}{3} \tilde{\gamma}^{ij}\partial_j\partial_k\beta^k 
 + \partial_k\tilde{\gamma}^{kj} \cdot \partial_j\beta^i
 - \frac{2}{3}\partial_k\tilde{\gamma}^{ki} \cdot \partial_j\beta^j\nonumber\\
 && - 2\tilde{A}^{ij}\partial_j\alpha 
 + 2\alpha\left[ (m-1)\partial_k\tilde{A}^{ki} - \frac{2m}{3}\tilde{D}^i K
    + m(\tilde{\Gamma}^i_{\; kl}\tilde{A}^{kl} + 6\tilde{A}^{ij}\partial_j\phi) \right]
    - S^i,
\label{Eq:BSSN8}
\end{eqnarray}
where we have decomposed the three metric and the extrinsic curvature according to
\begin{eqnarray}
\gamma_{ij} &=& e^{4\phi}\tilde{\gamma}_{ij}\; ,\\
K_{ij} &=& e^{4\phi}\left( \tilde{A}_{ij} + \frac{1}{3}\tilde{\gamma}_{ij} K \right),
\end{eqnarray}
$\tilde{\gamma}_{ij}$ having unit determinant and $\tilde{A}_{ij}$ being trace-free, and where we have introduced the operator $\hat{\partial}_0 = \partial_t - \beta^j\partial_j$. Here, all quantities with a tilde refer to the conformal three metric $\tilde{\gamma}_{ij}$, and the latter is used in order to raise and lower their indices. In particular, $\tilde{D}_i$ and $\tilde{\Gamma}^k_{\; ij}$ refer to the covariant derivative and the Christoffel symbols,
respectively, with respect to $\tilde{\gamma}_{ij}$. The expression $[ ... ]^{TF}$ denotes the trace-less part (with respect to the metric $\tilde{\gamma}_{ij}$) of the expression inside the parentheses, and
\begin{eqnarray}
\tilde{R}_{ij} &=& 
  - \frac{1}{2} \tilde{\gamma}^{kl}\partial_k\partial_l\tilde{\gamma}_{ij} 
  + \tilde{\gamma}_{k(i}\partial_{j)}\tilde{\Gamma}^k
  - \tilde{\Gamma}_{(ij)k}\partial_l\tilde{\gamma}^{kl} 
  + \tilde{\gamma}^{ls}\left( 2\tilde{\Gamma}^k_{\; l(i}\tilde{\Gamma}_{j)ks} 
  + \tilde{\Gamma}^k_{\; is}\tilde{\Gamma}_{klj} \right),
\nonumber\\
R^\phi_{ij} &=& -2\tilde{D}_i\tilde{D}_j\phi - 2\tilde{\gamma}_{ij} \tilde{D}^k\tilde{D}_k\phi
  + 4\tilde{D}_i\phi\, \tilde{D}_j\phi - 4\tilde{\gamma}_{ij}\tilde{D}^k\phi\, \tilde{D}_k\phi.
\nonumber
\end{eqnarray}
The parameter $m$, which was introduced in \cite{mAetal00}, controls how the momentum constraint is added to the evolution equations for the variable $\tilde{\Gamma}^i$. Also, $K_0$ is a smooth function of its argument, $f$, $G$ and $H$ are strictly positive and smooth functions of their arguments, and $\eta^i$ is a vector-valued function depending smoothly on its arguments. The choice
\begin{eqnarray}
&& m = 1,\qquad 
f(\alpha,\phi,x^\mu) = \frac{2}{\alpha}\; ,\qquad
K_0(x^\mu) = 0,\nonumber\\
&& G(\alpha,\phi,x^\mu) = \frac{3}{4\alpha^2}\; ,\qquad
H(\alpha,\phi,x^\mu) = e^{4\phi}\; ,\qquad
\eta^i(B^j,\alpha,x^\mu) = \eta B^i,
\nonumber
\end{eqnarray}
with $\eta$ a positive constant corresponds to the evolution system used in recent black hole simulations based on $1+\log$ slicing and the moving puncture technique (see, for instance, Ref. \cite{vMjBmKdC06}) or the turducken approach \cite{dBoSeSmTpDiHdP07,dBpDoSeSmT09}. The source terms $S$, $\hat{S}_{ij}$ and $S^i$ are defined in terms of the four Ricci tensor, $R^{(4)}_{ij}$, and the constraint variables\footnote{Notice that these constraint variables are related to variables $H$, $M_i$ and $C_\gamma^i$ defined in \cite{dBpDoSeSmT09} through ${\cal H} = H + e^{-4\phi}\partial_i C_\Gamma^i/2$, ${\cal M}_i = M_i$, ${\cal C}^i = C_\Gamma^i$.}
\begin{eqnarray}
{\cal H} &\equiv& \frac{1}{2}\left( \gamma^{ij}\tilde{R}_{ij} + \gamma^{ij} R^\phi_{ij}
  + \frac{2}{3} K^2 - \tilde{A}^{ij}\tilde{A}_{ij} \right),
\label{Eq:BSSNCons1}\\
{\cal M}_i &\equiv& \tilde{D}^j \tilde{A}_{ij} 
  - \frac{2}{3} \tilde{D}_i K + 6\tilde{A}_{ij} \tilde{D}^j\phi,
\label{Eq:BSSNCons2}\\
{\cal C}^i &\equiv& \tilde{\Gamma}^i + \partial_j\tilde{\gamma}^{ij},
\label{Eq:BSSNCons3}
\end{eqnarray}
as
\begin{eqnarray}
S &=& \gamma^{ij} R^{(4)}_{ij} - 2{\cal H} + e^{-4\phi}\partial_i {\cal C}^i,\\
\hat{S}_{ij} &=& \left[ R^{(4)}_{ij} 
 + \tilde{\gamma}_{k(i}\partial_{j)} {\cal C}^k \right]^{TF},\\
S^i &=& 2\alpha\, m\,\tilde{\gamma}^{ij}{\cal M}_j 
 - \hat{\partial}_0{\cal C}^i.
\end{eqnarray}
The Einstein field equations are equivalent to the evolution equations
(\ref{Eq:BSSN1}--\ref{Eq:BSSN8}) setting $S=-4\pi G_N(\rho+s)$, $\hat{S}_{ij}=8\pi G_N s_{ij}^{TF}$, $S^i = 16\pi G_N m\alpha\tilde{\gamma}^{ik} j_k$ and the constraints ${\cal H} = 8\pi G_N\rho$, ${\cal M}_i = 8\pi G_N j_i$ and ${\cal C}^i = 0$ with $G_N$ Newton's constant and where $\rho$, $j_i$ and $s_{ij}$ denote, respectively, the energy density, the momentum density and the stress tensor of the matter fields as measured by Eulerian observers, see \cite{Alcubierre-Book} for a definition.

Considering those readers who already want to see the end of the story, we directly present our main result, and describe their derivation in detail below. Our nine boundary conditions for the BSSN system with $m=1$, consistent with the constraint equations are given in Table~\ref{Tablita}.
\begin{table*}[!t]\centering
\caption{The nine boundary conditions for the BSSN variables.
Here, $n$ is the unit outward normal to the boundary which is tangent to the time foliation and normalized such that $n^i\,n_i=1$, where $n_i=\gamma\,_{ij}\,n^j$. Further, $\beta^n=\beta^i\,n_i$, $\kappa = 4G H/3$, $\partial_n=n^i\partial_i$ denotes the normal derivative and $\Pi^i_j=\delta^i_j - n^i\,n_j$ and ${P^{ij}}_{lm}=\Pi^i_l\,\Pi^j_m - \frac12\,\Pi^{ij}\,\Pi_{lm}$ denote the projectors onto tangential vectors and trace-less tangential tensors, respectively. We are also defining the quantities ${\bar{\cal E}}_{ij}=\tilde{R}_{ij}
+ {R^\phi}_{ij} + e^{4\,\phi}\,\left( \frac{1}{3}\, K\tilde{A}_{ij} -
\tilde{A}_{il}\,\tilde{A}^l_j\right) - 4\pi G_N s_{ij}$ and  ${\bar{\cal B}}_{kij}=
e^{4\,\phi}\,\left[\tilde{D}_k\,\tilde{A}_{ij} - 4\,\left(\tilde{D}_{(i}\,\phi\right)\tilde{A}_{j)k}\right]$ which determine the electric and magnetic parts of the Weyl tensor through $E_{ij} = {\bar{\cal E}}_{ij} - \frac{1}{3}\gamma_{ij}\gamma^{kl}{\bar{\cal E}}_{kl}$ and $B_{ij} = {\bar{\cal B}}_{kl(i}\varepsilon_{j)}{}^{kl}$, respectively, where $\varepsilon_{kij}$ denotes the volume form with respect to the three metric $\gamma_{ij}$. Finally, $G_{ij}$ is a given function on the boundary which determines the value of the Weyl scalar $\Psi_0$ at the boundary. The precise relation between $G_{ij}$ and $\Psi_0$ is the following: if $N = \alpha^{-1}(\partial_t - \beta^i\partial_i)$ denotes the future-directed unit normal to the time slices, we may construct a Newman-Penrose null tetrad $\{ l,k,m,\bar{m} \}$ at the boundary by defining the real null vectors $l:=2^{-1/2}(N + n)$, $k:=2^{-1/2}(N - n)$, and choosing $m$ to be a unit complex null vector orthogonal to $l$ and $k$. Then, $\Psi_0 = (E_{kl} - i B_{kl}) m^k m^l = G_{kl} m^k m^l$. For typical applications involving the modeling of isolated systems one may set $G_{ij}$ to zero or freeze its value to the one computed from the initial data.} 
\label{Tablita}
\begin{tabular}{c c c}
    \toprule
Description & Boundary Conditions  \\
\hline
Gauge condition on the lapse & $\partial_n\alpha=0$  \\
Gauge condition on the normal component of the shift &$\beta^n=0$ \\
Gauge condition on the tangential components of the shift
&$\Pi^i_j\,\left(\partial_t +
\frac{\sqrt{3\kappa}}2\partial_n\right)\beta^j=\frac{\kappa}{f-\kappa}
\Pi^i_j\,\tilde{\gamma}^{jk}\partial_k\alpha$ \\
Constraint preserving condition &$\tilde{D}^j \tilde{A}_{ij} 
  - \frac{2}{3} \tilde{D}_i K + 6\tilde{A}_{ij} \tilde{D}^j\phi = 8\pi G_N j_i$ \\
$\Psi_0$ specifying condition & ${P^{ij}}_{lm}{\bar{\cal E}}_{ij} + 
\left(n^k\,{P^{ij}}_{lm} - {n^i\,P^{kj}}_{lm} \right)\,{\bar{\cal
B}}_{kij}={P^{ij}}_{lm}\,G_{ij}$ \\
\hline
\end{tabular}
\end{table*}
Let us explain first why there should be nine boundary conditions. In \cite{hBoS04} the BSSN system was analyzed in terms of the eigenfields, and it was shown that under certain restrictions on $m$ and the functions $f$, $G$ and $H$ the Eqs.~(\ref{Eq:BSSN1}--\ref{Eq:BSSN8}) yield a strongly hyperbolic system giving rise to a well-posed time evolution. For $m=1$ the restrictions reduce to
\begin{equation}
4G H \neq 3f.
\end{equation}
The characteristic speeds with respect to the time evolution vector field $\partial_t$ are then
\begin{displaymath}
\beta^n, \qquad
\beta^n \pm \alpha\,\sqrt{f}, \qquad
\beta^n \pm \alpha\,\sqrt{G H},\qquad
\beta^n \pm \alpha\,\sqrt{\frac{4\,G H}3},\qquad
\beta^n \pm \alpha, 
\end{displaymath}
where $n$ refers to the unit outward normal to the boundary tangent to the time foliation.
This result is based on a first-order pseudo-differential reduction of the equations 
\cite{KreissOrtiz-2002,gNoOoR04} which does not introduce any artificial constraints.
Therefore, these speeds are intrinsic to the BSSN formulation. If $\partial_t$ is tangent to
the boundary, which is usually assumed in order to avoid the boundary from moving through the computational domain, the sign of these speeds determines the number of incoming characteristic fields and hence also the number of boundary conditions that must be specified. Namely, the number of boundary conditions is equal to the number of eigenfields with {\em positive} speed. Because of the presence of the modes with speed $\beta^n$ it is the sign of the normal component of the shift which determines the number of incoming fields at the boundary. The simplest way of controlling the number of boundary conditions throughout evolution is to demand that $\beta^n$ be zero at the boundary at all times. Otherwise, in order to gain control on the in- and out-going eigenfields, a careful study on the behavior of the shift vector would have to be implemented at each time step. Therefore, in our study, we take the simplest approach and set $\beta^n=0$ as one of our boundary conditions. The analysis in \cite{hBoS04} 
then reveals that there are precisely nine incoming eigenfields and thus, nine conditions 
have to be imposed at the boundary. If more conditions are given, the system will be overdetermined and there will not be solutions; if less conditions are given, the solutions will not be unique. 

As our analysis in the weak field regime shows, the nine boundary conditions are distributed as follows: there are four conditions that must be imposed for the gauge functions, namely the lapse and shift. One of these conditions sets the normal component of the shift to zero, as explained above. Geometrically, this implies that the boundary surface is orthogonal to the time foliation. Then, there is a Neumann boundary condition for the lapse. The other two gauge boundary conditions are Sommerfeld-like boundary conditions involving the tangential components of the shift and the tangential derivatives of the lapse. These conditions arise from the analysis of the characteristic structure of the gauge sector, and it would have been hard to guess them. As described in the next section, an alternative to these conditions is to directly specify the tangential components of the shift vector at the boundary. Next, there are three boundary conditions coming from the requirement, which is quite natural, of the momentum constraint being satisfied at the boundary. As we show in Sec.~\ref{Sec:CP}, these conditions imply constraint preservation without overdetermining the system. This means that the time evolution of initial data satisfying the constraints automatically satisfies the constraints on all time slices, and that small initial violations of the constraints which are usually present in numerical applications yield solutions to the evolution system where the growth of the constraint violations is controlled. In particular, we note that the Hamiltonian constraint must not be imposed additionally at the boundary, otherwise the resulting evolution system is overdetermined when small violations of the constraints are present. Finally, the last two boundary conditions are related to the actual gravitational degrees of freedom in the following sense: we are thinking about a problem which is localized and isolated and where the outer boundary is placed in the weak field and wave zone. In this case, the peeling behavior of the Weyl components \cite{rP65} can be considered to be valid, and the complex Weyl scalar $\Psi_0$ can be interpreted as approximately describing the incoming gravitational radiation. Therefore, it is natural to set $\Psi_0$ to zero or to its value computed from the initial data in order to minimize incoming gravitational radiation. While this condition only makes precise sense at null infinity, it has been successfully numerically implemented and tested for truncated domains with artificial boundaries, see for example \cite{oRlLmS07}. Estimates on the amount of spurious reflections introduced by this condition have also been derived in \cite{lBoS06,lBoS07}.

The boundary conditions in Table~\ref{Tablita} are extrapolated from the boundary conditions constructed in the next section for the linearized system. Some ideas about their numerical implementations are discussed in Sec.~\ref{Sec:Numer}.

\section{Analysis of the linearized system}
\label{Sec:LinSystem}

For the following, we restrict our analysis to the case $m=1$ since this is the common choice in numerical relativity applications.  Furthermore, we consider only small amplitude, high-frequency perturbations of smooth solutions, which intuitively, constitute the relevant limit for analyzing the continuous dependence of the solution on the initial data \cite{KL-Book}. In this limit only the principal part of the equations matters and the coefficient appearing in front of the derivative operators can be frozen to their value at an arbitrary point $p$. Therefore, the study simplifies to a linear evolution problem with constant coefficients on the half plane $\Sigma =  \{ (x,y,z)\in\Real^3 : x > 0 \}$. By rescaling and rotating the coordinates if necessary, one can bring the spacetime metric at $p$ in the following form (see \cite{mRoRoS07} for details),
\begin{displaymath}
\left. ds^2 \right|_p = -dt^2 + (dx + \beta^x dt)^2 + dy^2 + dz^2,
\end{displaymath}
where $\beta^x$ represents the normal component of the shift vector at $p$.\footnote{A
redefinition of $x$ is possible that makes this term vanish; however, this transformation does not leave the spacetime domain $M = [0,T] \times \Sigma$ invariant, but results in a domain with a "moving boundary".} As mentioned above, we set the normal component of the shift to zero since we are interested in controlling the number of ingoing fields at the boundary. Therefore, we freeze the coefficients in front of the derivative operators in the evolution equations (\ref{Eq:BSSN1}--\ref{Eq:BSSN8}) to the values
\begin{displaymath}
\alpha = 1, \qquad
\beta^k = 0,\qquad
\phi = 0, \qquad
\tilde{\gamma}_{ij} = \delta_{ij}\; .
\end{displaymath}
For the following, we use the standard operators from vector calculus
$\grad$, $\curl$, $\divr$ defined by
\begin{displaymath}
(\grad\phi)_i = \partial_i\phi, \qquad
(\curl X)_i = \varepsilon_{ikl}\partial^k X^l,\qquad
\divr X = \partial^k X_k,
\end{displaymath}
for scalar and vector fields $\phi$ and $X$, respectively. They satisfy the identities
\begin{eqnarray}
&& \curl\grad\phi = 0, \qquad
   \divr\curl X = 0, \qquad
   \divr\grad\phi = \Delta\phi,
\label{Eq:VectorId1}\\
&& \curl\curl X = -\Delta X + \grad\divr X,
\label{Eq:VectorId2}
\end{eqnarray}
where $\Delta = \partial^k\partial_k$ denotes the standard Laplacian.
We consider the following generalization from tensor calculus:
\begin{displaymath}
(\grad X)_{ij} := \partial_{(i} X_{j)} 
               - \frac{1}{3}\delta_{ij}\partial^k X_k, \qquad
(\curl T)_{ij} := \varepsilon_{kl(i}\partial^k T^l{}_{j)},\qquad
(\divr T)_j := \partial^i T_{ij},
\end{displaymath}
where $T$ is a symmetric, traceless tensor field. Notice that by
definition, $\grad X$ and $\curl T$ are symmetric, traceless tensor
fields. The following identities generalize the previous ones from
vector calculus:
\begin{eqnarray}
\curl\grad X &=& \frac{1}{2}\,\grad\curl X,
\label{Eq:TensorId1}\\
\divr\curl T &=& \frac{1}{2}\,\curl\divr T,
\label{Eq:TensorId2}\\
\divr\grad X &=& \frac{1}{2}\Delta X + \frac{1}{6}\,\grad\divr X,
\label{Eq:TensorId3}\\
\curl\curl T &=& -\Delta T + \frac{3}{2}\,\grad\divr T.
\label{Eq:TensorId4}
\end{eqnarray}
Notice that the identities (\ref{Eq:TensorId1},\ref{Eq:TensorId3})
imply that $\curl\grad\grad\phi = 0$ and $\divr\grad\grad\phi =
2\grad\Delta\Phi/3$. With this notation, the evolution equations
(\ref{Eq:BSSN1}--\ref{Eq:BSSN8}) in the high-frequency limit are
\begin{eqnarray}
\dot{\alpha} &=& -f_0 K,
\label{Eq:LinBSSN1}\\
\dot{K} &=& -\Delta\alpha,
\label{Eq:LinBSSN2}\\
\dot{\beta} &=& G_0 B,
\label{Eq:LinBSSN3}\\
\dot{B} &=& H_0\left( \Delta\beta + \frac{1}{3}\grad\divr\beta 
  - \frac{4}{3}\,\grad K \right),
\label{Eq:LinBSSN4}\\
\dot{\phi} &=& -\frac{1}{6} K + \frac{1}{6}\divr\beta,
\label{Eq:LinBSSN5}\\
\dot{\gamma} &=& -2A + 2\grad\beta,
\label{Eq:LinBSSN6}\\
\dot{A} &=& -\frac{1}{2}\Delta\gamma + \grad\Gamma - 2\grad\grad\phi
 - \grad\grad\alpha,
\label{Eq:LinBSSN7}\\
\dot{\Gamma} &=& \Delta\beta + \frac{1}{3}\grad\divr\beta 
  - \frac{4}{3}\,\grad K,
\label{Eq:LinBSSN8}
\end{eqnarray}
where $f_0$, $G_0$ and $H_0$ are the values of $f$, $G$ and $H$ frozen
at the point $p$, and where for notational simplicity we omit the
tildes over $\tilde{\gamma}$, $\tilde{A}$ and $\tilde{\Gamma}$ in what
follows. The constraints in the high-frequency limit are
\begin{eqnarray}
{\cal H} &\equiv& \frac{1}{2}\,\divr\Gamma - 4\Delta\phi = 0,
\label{Eq:LinHam}\\
{\cal M} &\equiv& \divr A - \frac{2}{3}\,\grad K = 0,
\label{Eq:LinMom}\\
{\cal C} &\equiv& \Gamma - \divr\gamma = 0.
\label{Eq:LinC}
\end{eqnarray}
The goal of this section is to construct boundary conditions at $x=0$ which ensure a
well-posed  Cauchy evolution for the above system, guaranteeing that the constraints propagate. Our construction is based on similar ideas described in \cite{oRoS05,gNoS06} and start with analyzing the propagation of the gauge-invariant quantities, namely the curvature variables consisting of the linearized Weyl tensor and the linearized constraint variables ${\cal H}$, ${\cal M}$ and ${\cal C}$ defined above.

\subsection{Propagation of the gauge-invariant quantities}
\label{SubSec:GaugeInvariant}

Here, we focus our attention on the evolution of the gauge-invariant quantities consisting of the constraint variables ${\cal H}$, ${\cal M}$ and ${\cal C}$ and the linearized Weyl
curvature tensor. The latter can be divided in electric and magnetic parts, which in terms of the BSSN variables, read
\begin{eqnarray}
{\cal E} &=& \dot{A} + \grad\grad\alpha
 = -\frac{1}{2}\Delta\gamma + \grad\Gamma - 2\grad\grad\phi,
\label{Eq:WeylE}\\
{\cal B} &=& \curl A.
\label{Eq:WeylB}
\end{eqnarray}
Using the identities (\ref{Eq:TensorId1}--\ref{Eq:TensorId4}) it is not difficult to see that
the evolution equations (\ref{Eq:LinBSSN2},\ref{Eq:LinBSSN5}--\ref{Eq:LinBSSN8}) imply the following FOSH system for these gauge-invariant quantities,
\begin{eqnarray}
\dot{\cal H} &=& 0,
\label{Eq:GaugeInv1}\\
\dot{\cal C} &=& 2{\cal M},
\label{Eq:GaugeInv2}\\
\dot{\cal M} &=& \divr{\cal E},
\label{Eq:GaugeInv3}\\
\dot{\cal E} &=& -\curl{\cal B} + \frac{3}{2}\,\grad{\cal M},
\label{Eq:GaugeInv4}\\
\dot{\cal B} &=& +\curl{\cal E}.
\label{Eq:GaugeInv5}
\end{eqnarray}
The following observation is important for the argument below: the gauge-invariant 
quantities ${\cal H}$, ${\cal M}$, ${\cal C}$, ${\cal E}$ and ${\cal B}$ are not independent of each other. Indeed, one has $2\divr{\cal B} = 2\divr\curl A = \curl\divr A = \curl{\cal M}$, resulting in the "super-constraint" $P:=2\divr{\cal B} - \curl{\cal M} = 0$. Also, one has
$2\divr{\cal E} = -\Delta\divr\gamma - 8\grad\Delta\phi/3 + \Delta\Gamma + \grad\divr\Gamma/3 = \Delta{\cal C} - 2\grad{\cal H}/3$, resulting in the super-constraint $Q:=2\divr{\cal E} - \Delta C - 2\grad{\cal H}/3 = 0$. Therefore, one can modify the system above by adding multiples of the constraint $P = 0$ and $Q = 0$ to the equations.

As an example, we may use either one of the super-constraints $P=0$ or $Q=0$ in order to derive the following decoupled system for the constraint variables ${\cal H}$, ${\cal M}$ and
${\cal C}$:
\begin{eqnarray}
\dot{\cal H} &=& 0,
\label{Eq:Constraint1}\\
\dot{\cal C} &=& 2{\cal M},
\label{Eq:Constraint2}\\
\ddot{\cal M} &=& \Delta{\cal M}.
\label{Eq:Constraint3}
\end{eqnarray}
Our boundary conditions must ensure that these constraints propagate, that is, initial 
data satisfying ${\cal H} = 0$, ${\cal M} = 0$, ${\cal C} = 0$ must lead to solutions
satisfying these constraints everywhere on the spacetime domain. A simple condition which implies this property is the imposition of a homogeneous Dirichlet condition on the momentum constraint variable ${\cal M}$,
\begin{equation}
{\cal M} \hateq 0,
\label{Eq:MomentumCondition}
\end{equation}
where here and in the following, the symbol $\hateq$ refers to equality at the boundary 
surface $x=0$. In the next section, we show that this condition also leads to
constraint-propagation for the full nonlinear BSSN equations.

Now we go back to the full propagation system for the gauge-invariant quantities 
(\ref{Eq:GaugeInv1}--\ref{Eq:GaugeInv5}), and try to specify boundary conditions for this
system which incorporates the three constraint-preserving boundary conditions
(\ref{Eq:MomentumCondition}). In order to determine which boundary conditions may be specified for this system, we analyze the corresponding principal symbol and its characteristics (see the appendix). The non-trivial part of the principal symbol is
\begin{equation}
{\bf A}(n)\left( \begin{array}{l} {\cal M} \\ {\cal E} \\  {\cal B} \end{array} \right)
 = \left( \begin{array}{l}  
 n\cdot{\cal E} \\
 -n \wedge {\cal B} + \frac{3}{2}\,n \otimes {\cal M} \\
  +n \wedge {\cal E} \end{array} \right),
\end{equation}
where $n$ is a one-form and where we use the notation $(n\wedge {\cal B})_{ij} := 
\varepsilon_{kl(i} n^k {\cal B}^l{}_{j)}$, $(n\otimes {\cal M})_{ij} := n_{(i} {\cal M}_{j)} -
\frac{1}{3}\delta_{ij} n^k M_k$, $(n\cdot
{\cal E})_j := n^i{\cal E}_{ij}$. Decomposing
\begin{eqnarray}
{\cal M} &=& {\cal M}_{||}\, n + {\cal M}_{\perp},
\nonumber\\
{\cal E} &=& \frac{3}{2} {\cal E}_{||||}\, n\otimes n 
 + 2n\otimes {\cal E}_{||\perp} + {\cal E}_{\perp\perp},
\nonumber\\
{\cal B} &=& \frac{3}{2} {\cal B}_{||||}\, n\otimes n 
 + 2n\otimes {\cal B}_{||\perp} + {\cal B}_{\perp\perp},
\nonumber
\end{eqnarray}
into pieces parallel and orthogonal to $n$, the eigenvalue problem ${\bf A}(n) u = \lambda u$ yields the three scalar equations
\begin{equation}
\lambda{\cal M}_{||} = {\cal E}_{||||},\qquad
\lambda{\cal E}_{||||} = {\cal M}_{||},\qquad
\lambda{\cal B}_{||||} = 0,
\label{Eq:GaugeInvScalarBlock}
\end{equation}
the three vector equations
\begin{equation}
\lambda{\cal M}_{\perp} = {\cal E}_{||\perp},\qquad
\lambda{\cal E}_{||\perp} = -\frac{1}{2}\, n\wedge {\cal B}_{||\perp}
 + \frac{3}{4}\,{\cal M}_{\perp},\qquad
\lambda{\cal B}_{||\perp} = \frac{1}{2}\, n\wedge {\cal E}_{||\perp},
\label{Eq:GaugeInvVectorBlock}
\end{equation}
and the two tensor equations
\begin{equation}
\lambda{\cal E}_{\perp\perp} = -n\wedge {\cal B}_{\perp\perp},\qquad
\lambda{\cal B}_{\perp\perp} =  n\wedge {\cal E}_{\perp\perp}.
\label{Eq:GaugeInvTensorBlock}
\end{equation}
From this we obtain the following characteristic fields with zero speeds
\begin{displaymath}
{\cal B}_{||||},\qquad
Z^{(0)}_\perp = {\cal B}_{||\perp} - \frac{1}{2}\, n \wedge {\cal M}_\perp,
\end{displaymath}
and the ones with nonzero speeds,
\begin{eqnarray}
V^{(\pm 1)} &=& {\cal E}_{||||} \pm {\cal M}_{||},
\nonumber\\
V^{(\pm 1)}_\perp &=& 4{\cal E}_{||\perp} \mp 2n\wedge {\cal B}_{||\perp}
 \pm 3{\cal M}_\perp,
\nonumber\\
V^{(\pm 1)}_{\perp\perp} &=& {\cal E}_{\perp\perp} \mp n\wedge {\cal B}_{\perp\perp},
\nonumber
\end{eqnarray}
with corresponding characteristic speeds $\lambda$ indicated by the superscripts $(\pm 1)$ and $(0)$. Maximal dissipative boundary conditions have the form of a linear coupling between the in- and outgoing fields corresponding to the outward unit one-form normal to the boundary, i.e. $n = -\partial_x$ (see the appendix). In our case, this means that we may specify conditions of the following form,
\begin{equation}
V^{(+1)} \hateq c_0 V^{(-1)} + G,\qquad
V^{(+1)}_\perp \hateq c_1 V^{(-1)}_\perp + G_\perp,\qquad
V^{(+1)}_{\perp\perp} \hateq c_2 V^{(-1)}_{\perp\perp} + G_{\perp\perp},
\label{Eq:MaxDissCurv}
\end{equation}
with coupling constants $c_0$, $c_1$ and $c_2$ and boundary data $G$, $G_\perp$ and
$G_{\perp\perp}$. In order to fix the coupling constants $c_0$, $c_1$ and $c_2$ we first remark that the fields $V^{(\pm 1)}_{\perp\perp}$ are related to the linearized Weyl scalars 
\begin{displaymath}
\Psi_0 = C_{\alpha\beta\gamma\delta} l^\alpha m^\beta l^\gamma m^\delta, \qquad
\Psi_4 = C_{\alpha\beta\gamma\delta} k^\alpha\bar{m}^\beta k^\gamma\bar{m}^\delta
\end{displaymath}
in the following way: the Newman-Penrose null tetrad $\{ l^\alpha, k^\alpha, m^\alpha,
\bar{m}^\alpha \}$ is defined by the time-evolution vector field $\partial_t$ and the normal to the boundary $n =
-\partial_x$ according to
\begin{displaymath}
l = \frac{1}{\sqrt{2}}\left( \partial_t + n \right),\qquad
k =  \frac{1}{\sqrt{2}}\left( \partial_t - n \right),
\end{displaymath}
up to a rotation $m\to e^{i\varphi} m$. Then, we have,
\begin{displaymath}
V^{(+1)}_{\perp\perp} 
 = (\Psi_0\bar{m}\otimes\bar{m} + \bar{\Psi}_0 m\otimes m),
\qquad
V^{(-1)}_{\perp\perp} 
 = (\Psi_4 m\otimes m + \bar{\Psi}_4\bar{m}\otimes\bar{m}).
\end{displaymath}
Therefore, choosing $c_2=0$ in Eq. (\ref{Eq:MaxDissCurv}) we obtain the boundary condition
\begin{equation}
V^{(+1)}_{\perp\perp} \hateq G_{\perp\perp},
\label{Eq:Psi0Freezing}
\end{equation}
which is equivalent to specifying the Weyl scalar $\Psi_0$ at the boundary. In particular, 
one might choose $G_{\perp\perp} \hateq \left. V^{(+1)}_{\perp\perp} \right|_{t=0}$ which freezes $\Psi_0$ to its initial value. This condition has been shown to yield a reflection coefficient that decays as fast as $(k R)^{-4}$ for monochromatic gravitational radiation with wave number $k$ and a spherical outer boundary of radius $R$ \cite{lBoS06,lBoS07}. It has also been tested numerically \cite{oRlLmS07} for gravitational waves propagating about a Schwarzschild black hole and shown to outperform other currently used boundary conditions. 

Next, choosing $c_0=1$ and $G=0$ in Eq. (\ref{Eq:MaxDissCurv}), is equivalent to setting the constraint ${\cal M}_{||}$ to zero at the boundary. On the other hand, the form of $V^{(\pm 1)}_\perp$ does not allow one to set the orthogonal component, ${\cal M}_\perp$, of the momentum constraint variable to zero at the boundary.\footnote{The boundary condition ${\cal M}_\perp \hateq 0$ is equivalent to $V^{(+1)}_\perp \hateq V^{(-1)}_\perp - 4 n\wedge Z^{(0)}_\perp$, and involves a zero speed field.} For this reason, we perform a slight modification to the propagation system (\ref{Eq:GaugeInv1}--\ref{Eq:GaugeInv5}) by using the super-constraint $P=2\divr{\cal B} - \curl{\cal M} = 0$. Namely, we replace Eq. (\ref{Eq:GaugeInv4}) by
\begin{equation}
\dot{\cal E} = -\curl{\cal B} + \frac{3}{2}\,\grad{\cal M}
 + n \otimes \left[ n\wedge \left( \divr{\cal B} - \frac{1}{2}\curl{\cal M} 
 \right) \right],
\label{Eq:GaugeInv4bis}
\end{equation}
where $n = -\partial_x$ is the outward unit one-form normal to the boundary.\footnote{This modification is partially motivated by the "boundary adapted system" in Ref. \cite{hFgN99}.} With this modification, the only change in the eigenvalue problem ${\bf A}(n) u = u$ is the middle vector equation in Eq.~(\ref{Eq:GaugeInvVectorBlock}) which now simplifies to $\lambda{\cal E}_{||\perp} = {\cal M}_\perp$. Therefore, the characteristic fields remain unchanged except for the fields $V^{(\pm 1)}_\perp$ which have to be replaced by
\begin{equation}
V^{(\pm 1)}_\perp = {\cal E}_{||\perp} \pm {\cal M}_\perp\; .
\end{equation}
With this modification, it is now possible to impose the momentum constraint at the boundary by choosing $c_0=c_1=1$ and $G=0$ and $G_{\perp}=0$ in Eq.~(\ref{Eq:MaxDissCurv}). However, it remains to prove that these boundary conditions, together with the condition (\ref{Eq:Psi0Freezing}) on the Weyl scalar $\Psi_0$ are maximal dissipative and that the modified evolution system
(\ref{Eq:GaugeInv1},\ref{Eq:GaugeInv2},\ref{Eq:GaugeInv3},\ref{Eq:GaugeInv4bis},
\ref{Eq:GaugeInv5}) is still symmetric hyperbolic. For this, it is convenient to replace ${\cal B}$ by the new variable ${\cal K}$ defined by
\begin{displaymath}
{\cal K} := {\cal B} - n \otimes (n\wedge {\cal M}).
\end{displaymath}
In order to write down the principal symbol of the resulting evolution equations, we choose 
standard Cartesian coordinates $x,y,z$ on $\Sigma$ such that $n = -\partial_x$. The principal symbol with respect to an arbitrary one-form $m = m_x dx + m_A dx^A$ then reads
\begin{equation}
{\bf A}(m)\left( \begin{array}{l}
{\cal E}_{xx} \\ {\cal K}_{xx} \\ {\cal M}_x \\ {\cal E}_{xB} \\ {\cal K}_{xB} \\
{\cal M}_B \\ \hat{\cal E}_{AB} \\ \hat{\cal K}_{AB} \end{array} \right)
 = \left( \begin{array}{l}
  -\varepsilon^{AB} m_A {\cal K}_{Bx} + m_x {\cal M}_x - m^A {\cal M}_A\\
 +\varepsilon^{AB} m_A {\cal E}_{Bx}\\
 m_x {\cal E}_{xx} + m^A {\cal E}_{Ax}\\
 -\varepsilon^{CD} m_C\hat{\cal K}_{DB} 
 - \frac{1}{2}\varepsilon_B{}^C m_C {\cal K}_{xx}
 + m_x {\cal M}_B + \frac{1}{2} m_B{\cal M}_x\\
 +\varepsilon^{CD} m_C\hat{\cal E}_{DB} 
 + \frac{1}{2}\varepsilon_B{}^C m_C {\cal E}_{xx}\\
m_x {\cal E}_{xB} + m^A\hat{\cal E}_{AB} - \frac{1}{2} m_B {\cal E}_{xx}\\
 -m_x\varepsilon_{C(A}\hat{\cal K}^C{}_{B)} + \varepsilon_{C(A} m^C {\cal K}_{B)x} 
- \frac12\delta_{AB}\varepsilon^{CD} m_C {\cal K}_{Dx}
+ m_{(A}{\cal M}_{B)} - \frac12\delta_{AB} m^C{\cal M}_C\\
+m_x\varepsilon_{C(A}\hat{\cal E}^C{}_{B)} - \varepsilon_{C(A} m^C {\cal E}_{B)x} 
+ \frac12\delta_{AB}\varepsilon^{CD} m_C {\cal E}_{Dx}
\end{array} \right),
\end{equation}
where the indices $A,B,C,D$ refer to the coordinates $y$ and $z$, and where we have defined  $\hat{\cal E}_{AB} := {\cal E}_{AB} - \frac12\delta_{AB}\delta^{CD}{\cal E}_{CD}$ and analogous for $\hat{\cal K}_{AB}$. It is straightforward to verify that this symbol is
symmetric with respect to the symmetrizer ${\bf H}$ defined by
\begin{eqnarray}
U^T {\bf H} U &=& {\cal E}_{xx}^2 + {\cal K}_{xx}^2 + {\cal M}_x^2
 + 2\delta^{AB}{\cal E}_{xA}{\cal E}_{xB} 
 + 2\delta^{AB}{\cal K}_{xA}{\cal K}_{xB}
 + 2\delta^{AB}{\cal M}_A{\cal M}_B
\nonumber\\
&+& 2\delta^{AC}\delta^{BD}\hat{\cal E}_{AB}\hat{\cal E}_{CD}
 + 2\delta^{AC}\delta^{BD}\hat{\cal K}_{AB}\hat{\cal K}_{CD},
\nonumber
\end{eqnarray}
where $U = ({\cal E}_{xx},{\cal K}_{xx},{\cal M}_x,{\cal E}_{xB},{\cal K}_{xB},{\cal M}_B,
\hat{\cal E}_{AB}, \hat{\cal K}_{AB})^T$, that is, ${\bf H}{\bf A}(m) = {\bf A}(m)^T {\bf H}$ for all one-forms $m$. In a coordinate-independent notation, this reads
\begin{displaymath}
U^T {\bf H} U = |{\cal E}_{||||}|^2 + |{\cal K}_{||||}|^2 + |{\cal M}_{||}|^2
 + 2|{\cal E}_{||\perp}|^2 + 2|{\cal K}_{||\perp}|^2 + 2|{\cal M}_\perp|^2
 + 2|{\cal E}_{\perp\perp}|^2 + 2|{\cal K}_{\perp\perp}|^2.
 \nonumber
\end{displaymath}
In particular, we have,
\begin{eqnarray}
U^T {\bf H}{\bf A}(n) U 
&=&  -2{\cal E}_{xx} {\cal M}_x - 4{\cal E}_{xA}{\cal M}^A 
  - 4\hat{\cal E}^{AB}\varepsilon_{AC}\hat{\cal K}^C{}_B
\nonumber\\
&=& \frac{1}{2}\left( |V^{(+1)}|^2 -  |V^{(-1)}|^2 \right)
 + \left( |V^{(+1)}_\perp|^2 -  |V^{(-1)}_\perp|^2 \right)
 + \left( |V^{(+1)}_{\perp\perp}|^2 -  |V^{(-1)}_{\perp\perp}|^2 \right).
\nonumber
\end{eqnarray}
For our choice of boundary conditions, $V^{(+1)} \hateq V^{(-1)}$, $V^{(+1)}_\perp \hateq
V^{(-1)}_\perp$ and $V^{(+1)}_{\perp\perp} \hateq G_{\perp\perp}$ we obtain
\begin{displaymath}
U^T {\bf H}{\bf A}(n) U = |G_{\perp\perp}|^2 -  |V^{(-1)}_{\perp\perp}|^2
\leq  |G_{\perp\perp}|^2.
\end{displaymath}
From this, we see that our boundary conditions are maximal dissipative.

We may summarize the results of this subsection in the following
\begin{lemma}
\label{Lem:GaugeInv}
The IBVP on the spacetime domain $M:=[0,\infty)\times\Sigma$ consisting of the evolution equations (\ref{Eq:GaugeInv1},\ref{Eq:GaugeInv2},\ref{Eq:GaugeInv3},\ref{Eq:GaugeInv4bis},\ref{Eq:GaugeInv5}), initial data for ${\cal H}$, ${\cal C}$, ${\cal M}$, ${\cal E}$ and ${\cal B}$ in $L^2(\Sigma)$ and the boundary conditions
\begin{displaymath}
{\cal M} \hateq 0,\qquad
{\cal E}_{\perp\perp} - n\wedge {\cal B}_{\perp\perp} \hateq G_{\perp\perp},
\end{displaymath}
where $n = -\partial_x$ is the unit outward vector to $\partial\Sigma$ and $G_{\perp\perp}$ lies in $L^2({\cal T})$, ${\cal T}:= [0,\infty)\times\partial\Sigma$,  is well posed.
\end{lemma}

The first boundary condition guarantees constraint-preservation, and the second condition 
specifies data to the Weyl scalar $\Psi_0$. In particular, we obtain an $L^2$ estimate for the gauge-invariant variables ${\cal H}$, ${\cal C}$, ${\cal M}$, ${\cal E}$ and ${\cal B}$. In view of Eqs. (\ref{Eq:LinMom},\ref{Eq:WeylE},\ref{Eq:WeylB}) this yields $L^2$ estimates for $\dot{A}$, $\curl A$ and $\divr A$ provided we have appropriate estimates for the lapse, $\alpha$, and the trace of the extrinsic curvature, $K$ (see the next subsection). We might ask whether or not the $L^2$ bounds for $\curl A$ and $\divr A$ are sufficient to bound the $L^2$ norm of the full gradient of $A$. On $\Real^3$, it can be proven that an $L^2$ bound on $\curl A$ and $\divr A$ implies an $L^2$ bound on $\grad A$. However, this is not true in general on the half-plane $\Sigma$ as the following example shows: Let $\chi$ be an arbitrary {\em harmonic} function on $\Sigma$ which decays exponentially to zero as $|x|\to\infty$ and let $A := \grad\grad \chi$. Then, $\curl A = 0$ and $\divr A = 2\grad\Delta\chi/3 = 0$, but $A\neq 0$ unless boundary conditions at $x=0$ force $\chi$ to be linear.

\subsection{Propagation of lapse and shift}
\label{SubSec:Gauge}

Next, we analyze the gauge sector, that is, the propagation of lapse and shift described in 
the small amplitude, high-frequency limit by the evolution Eqs.
(\ref{Eq:LinBSSN1}--\ref{Eq:LinBSSN4}). It is useful to introduce the quantities $a :=
-f_0^{-1}\grad\alpha$, $D:=G_0^{-1}\divr\beta$, $R:= G_0^{-1}\curl\beta$ in terms of which
these equations yield the following first-order system
\begin{eqnarray}
\dot{\alpha} &=& -f_0\, K,
\label{Eq:Gauge1}\\
\dot{\beta} &=& G_0 B,
\label{Eq:Gauge2}\\
\dot{K} &=& f_0\,\divr a,
\label{Eq:Gauge3}\\
\dot{a} &=& \grad K,
\label{Eq:Gauge4}\\
\dot{B} &=& \kappa_0\left( -\frac{3}{4}\curl R + \grad D - \frac{1}{G_0}\grad K \right),
\label{Eq:Gauge5}\\
\dot{R} &=& \curl B,
\label{Eq:Gauge6}\\
\dot{D} &=& \divr B,
\label{Eq:Gauge7}
\end{eqnarray}
where $\kappa_0 := 4G_0 H_0/3$ and where we have used the identity (\ref{Eq:VectorId2}). As in 
the previous subsection, this system is subject to super-constraints, which are
\begin{eqnarray}
{\cal C}_a &:=& a + f_0^{-1}\grad\alpha = 0,
\nonumber\\
{\cal C}_D &:=& D - G_0^{-1}\divr\beta = 0,
\nonumber\\
{\cal C}_R &:=& R - G_0^{-1}\curl\beta = 0.
\nonumber
\end{eqnarray}
As we show now, this system can be cast into FOSH form provided the condition
\begin{equation}
\kappa_0 \neq f_0
\end{equation}
holds. In order to see this, we rewrite the last three equations in terms of the new fields
\begin{displaymath}
C := B + \frac{\kappa_0 f_0}{G_0(f_0 - \kappa_0)} a,\qquad
F := D + \frac{\kappa_0}{G_0(f_0 - \kappa_0)} K,
\end{displaymath}
and use the constraint ${\cal C}_a = 0$, obtaining
\begin{eqnarray}
\dot{C} &=& \kappa_0\left( -\frac{3}{4}\curl R + \grad F \right),
\label{Eq:Gauge5Bis}\\
\dot{R} &=& \curl C,
\label{Eq:Gauge6Bis}\\
\dot{F} &=& \divr C.
\label{Eq:Gauge7Bis}
\end{eqnarray}
The system (\ref{Eq:Gauge1},\ref{Eq:Gauge2},\ref{Eq:Gauge3},\ref{Eq:Gauge4},\ref{Eq:Gauge5Bis},\ref{Eq:Gauge6Bis},\ref{Eq:Gauge7Bis}) is symmetric hyperbolic;
a symmetrizer $\hat{\bf H}$ for the principal symbol $\hat{\bf A}(n)$ is given by the quadratic form
\begin{displaymath}
v^T\hat{\bf H} v 
 = \alpha^2 + |\beta|^2 
 + K^2 + f_0|a|^2 + |C|^2 + \frac{3}{4}\kappa_0|R|^2 + \kappa_0 F^2,
\end{displaymath}
where $v = (\alpha,\beta,K,a,C,R,F)^T$. In terms of the characteristic fields
\begin{displaymath}
Y^{(\pm \sqrt{f_0})} := K \pm \sqrt{f_0}\, a_{||},\qquad
W^{(\pm \sqrt{\kappa_0})}_{||} := C_{||} \pm \sqrt{\kappa_0}\, F, \qquad
W^{(\pm \sqrt{G_0 H_0})}_\perp := C_\perp \mp \frac{1}{2}\sqrt{3\kappa_0}\, n \wedge R,
\end{displaymath}
we have
\begin{eqnarray}
v^T\hat{\bf H}\hat{\bf A}(n) v 
&=&  2f_0\, K a_{||} + 2\kappa_0 F C_{||} + \frac{3}{2}\kappa_0(C \wedge R)_{||}
\nonumber\\
&=& \frac{\sqrt{f_0}}{2}\left[ |Y^{(+\sqrt{f_0})}|^2 - |Y^{(-\sqrt{f_0})}|^2 \right]
 + \frac{\sqrt{\kappa_0}}{2}
     \left[ |W^{(+\sqrt{\kappa_0})}_{||}|^2 - |W^{(-\sqrt{\kappa_0})}_{||}|^2 \right]
\nonumber\\
&+& \frac{\sqrt{3\kappa_0}}{4}
     \left[ |W^{(+\sqrt{G_0 H_0})}_\perp|^2 - |W^{(+\sqrt{G_0 H_0})}_\perp|^2 \right].
 \nonumber
\end{eqnarray}
Therefore, we are allowed to impose the following boundary conditions at $x=0$,
\begin{equation}
Y^{(+\sqrt{f_0})} \hateq a_0 Y^{(-\sqrt{f_0})} + g,\qquad
W^{(+\sqrt{\kappa_0})}_{||} \hateq a_1 W^{(-\sqrt{\kappa_0})}_{||} + g_{||},\qquad
W^{(+\sqrt{G_0 H_0})}_{\perp} \hateq a_2 W^{(-\sqrt{G_0 H_0})}_{\perp} + g_{\perp},
\label{Eq:MaxDissGauge}
\end{equation}
with coupling constants $a_0$, $a_1$ and $a_2$ smaller than or equal to one in magnitude and boundary data $g_0$, $g_{||}$ and $g_{\perp}$, where we recall that $||$ and $\perp$ refer to components parallel and orthogonal to $n = -\partial_x$, respectively. In terms of lapse and shift, this is equivalent to
\begin{eqnarray}
&& (a_0-1)\dot{\alpha} \hateq \sqrt{f_0}(a_0+1)\partial_{||}\alpha + f_0 g,\\
&& (a_1-1)
 \left( \dot{\beta}_{||} - \frac{\kappa_0}{f_0 - \kappa_0}\partial_{||}\alpha \right)
\hateq \sqrt{\kappa_0}(a_1+1)\left( \divr\beta - \frac{\kappa_0}{f_0 -
\kappa_0}\frac{\dot{\alpha}}{f_0} \right) 
- G_0 g_{||},\\
&& (a_2-1)
 \left( \dot{\beta}_{\perp} - \frac{\kappa_0}{f_0 - \kappa_0}\partial_{\perp}\alpha \right)
\hateq \frac{\sqrt{3\kappa_0}}{2}(a_2+1)\left( \partial_{||}\beta_\perp - \partial_\perp\beta_{||} \right) 
  - G_0 g_{\perp}.
\end{eqnarray}
As mentioned in the introduction, we fix the normal component, $\beta_{||}$, of
the shift to zero. This requires that $a_0=-a_1=1$ and $g=g_{||}=0$. Choosing also $a_2=0$,\footnote{Here, different choices for $a_2$ are also possible. The choice $a_2=0$ is motivated by requiring a Sommerfeld-like boundary condition for the orthogonal part, $\beta_{\perp}$, of the shift vector. A natural alternative is $a_2=1$ which (together with the condition $\beta_{||}\hateq 0$) controls the tangential components of the shift at the boundary.} we obtain the following boundary conditions in the gauge
sector,
\begin{equation}
\partial_{||}\alpha \hateq 0, \qquad
\beta_{||} \hateq 0, \qquad
\left( \partial_t + \frac{\sqrt{3\kappa_0}}{2}\,\partial_{||} \right)\beta_{\perp} 
 \hateq \frac{\kappa_0}{f_0 - \kappa_0}\,\partial_{\perp}\alpha + G_0 g_\perp.
\label{Eq:LinGaugeBC}
\end{equation}

Summarizing, we have
\begin{lemma}
\label{Lem:Gauge}
The IBVP consisting of the evolution equations
(\ref{Eq:Gauge1},\ref{Eq:Gauge2},\ref{Eq:Gauge3},\ref{Eq:Gauge4},\ref{Eq:Gauge5Bis},\ref{Eq:Gauge6Bis},\ref{Eq:Gauge7Bis}), initial data for $\alpha$, $\beta$, $K$, $a$, $C$, $R$ and $F$ in $L^2(\Sigma)$ and the boundary conditions (\ref{Eq:LinGaugeBC}) with boundary data $g_\perp\in L^2({\cal T})$ is well posed.

Furthermore, the solution satisfies the constraints $\grad\alpha = -f_0 a$, $\divr\beta = G_0 F - \kappa_0 K/(f_0 - \kappa_0)$, $\curl\beta = G_0 R$ if the initial data satisfy these constraints.
\end{lemma}

\proof The first part follows from the considerations above, since we have a FOSH system with maximal dissipative boundary conditions. For the second part, we notice that the FOSH system implies the following evolution equations for the constraint variables ${\cal C}_a$, ${\cal C}_D$ and ${\cal C}_R$:
\begin{displaymath}
\dot{\cal C}_a = 0, \qquad
\dot{\cal C}_D = 0,\qquad
\dot{\cal C}_R = \frac{\kappa_0 f_0}{G_0(f_0-\kappa_0)}\curl{\cal C}_a.
\end{displaymath}
\qed

In particular, we obtain $L^2$ estimates for lapse and shift and their derivatives
$\dot{\alpha} = -f_0 K$, $\dot{\beta} = G_0 B$, $\grad\alpha = -f_0 a$, $\curl\beta = G_0 R$, $\divr\beta = G_0 D$. With the boundary condition $\beta_{||} \hateq 0$ this also implies an $L^2$ estimate for the symmetric and traceless gradient, $\grad\beta$, of the shift. This follows from the integral identity
\begin{displaymath}
\int\limits_{\Sigma} |\grad\beta|^2 d^3 x
 = \int\limits_{\Sigma} 
   \left( \frac{1}{2} |\curl\beta|^2 + \frac{2}{3} |\divr\beta|^2 \right) d^3 x
 - 2\int\limits_{\partial\Sigma} \beta_{||}\divr\beta_{\perp} dy dz
\end{displaymath}
which can be proven using integration by parts. Therefore, we obtain $L^2$ estimates for lapse and shift and all their first-order derivatives. Their second derivatives can be estimated by first taking time- and tangential derivatives of the evolution equations and boundary conditions and repeating the above analysis to obtain an $L^2$ bound for
$\dot{\alpha}$ and $\dot{\beta}$, $\partial_\perp\alpha$, $\partial_\perp\beta$ and their first derivatives and then estimating the second normal derivatives $\partial_{||}^2\alpha$ and $\partial_{||}^2\beta$ using the equations $\ddot{\alpha} = f_0\Delta\alpha$ and $\ddot{\beta} = \kappa_0( 3\Delta\beta + \grad\divr\beta - 4\grad K)/4$.

\subsection{Reconstructing the metric and connection variables}
\label{SubSec:Metric}

Finally, we discuss how the metric variables $\gamma$, $\phi$ and the connection variables $A$ and $\Gamma$ may be reconstructed from the gauge-invariant quantities ${\cal H}$, ${\cal C}$, ${\cal M}$, ${\cal E}$ and ${\cal B}$ and lapse and shift. First, $\phi$ and $\Gamma$ may be obtained after integrating the linearized BSSN Eqs.~(\ref{Eq:LinBSSN5}) and (\ref{Eq:LinBSSN8}), respectively. Then, the trace-free part of the extrinsic curvature is obtained by integrating the equation $\dot{A} = {\cal E} - \grad\grad\alpha$ in time. Then, $\gamma$ is obtained after integrating Eq. (\ref{Eq:LinBSSN6}) in time. It remains to show that the resulting fields yield a solution to the linearized BSSN system equations with the required boundary conditions. We state our main result in

\begin{theorem}[Well-posedness for the linearized IBVP]
\label{Thm:Main}
Let $\Sigma =  \{ (x,y,z)\in\Real^3 : x > 0 \}$ be the half plane, 
$M := [0,\infty) \times \Sigma$ spacetime and ${\cal T} := [0,\infty)\times \partial\Sigma$ the time-like boundary. Denote by $n = -\partial_x$ the unit outward vector to $\partial\Sigma$. Let $\kappa_0 = 4G_0 H_0/3$ and suppose the condition $\kappa_0 \neq f_0$ holds.

Given initial data for the fields $\alpha$, $K$, $\beta$, $B$, $\phi$, $\gamma$, $A$, $\Gamma$ in $C^\infty_0(\Sigma)$ at $t=0$, and given boundary data $G_{\perp\perp}$ and $g_\perp$ in $C^\infty_0({\cal T})$, there exists a unique smooth solution of the linearized BSSN system (\ref{Eq:LinBSSN1}--\ref{Eq:LinBSSN8}) on $M$ satisfying the boundary conditions
\begin{eqnarray}
\divr A - \frac{2}{3}\grad K &\hateq& 0,
\label{Eq:LinMomBC}\\
{\cal E}_{\perp\perp} - n\wedge {\cal B}_{\perp\perp} &\hateq& G_{\perp\perp},
\label{Eq:LinPsi0BC}\\
\partial_{||}\alpha &\hateq& 0,
\label{Eq:LinLapseBC}\\
\beta_{||} &\hateq& 0,
\label{Eq:LinBetaNBC}\\
\left( \partial_t + \frac{\sqrt{3\kappa_0}}{2}\,\partial_{||} \right)\beta_{\perp} 
 &\hateq& \frac{\kappa_0}{f_0 - \kappa_0}\,\partial_{\perp}\alpha + G_0 g_\perp,
\label{Eq:LinBetaOBC}
\end{eqnarray}
where ${\cal E} = -\frac{1}{2}\Delta\gamma + \grad\Gamma - 2\grad\grad\phi$ and ${\cal B} = \curl A$ are the electric and magnetic part, respectively, of the linearized Weyl tensor.

Furthermore, if the initial data satisfy the linearized constraints ${\cal H} = 0$, ${\cal M}
= 0$ and $C = 0$ defined in (\ref{Eq:LinHam},\ref{Eq:LinMom},\ref{Eq:LinC}), then the solution automatically satisfies these constraints on $M$.
\end{theorem}

\proof 
{\em Existence}: In a first step, we define the gauge-invariant quantities ${\cal H}$, ${\cal
M}$, ${\cal C}$, ${\cal E}$, ${\cal B}$ at $t=0$ by using the Eqs.~(\ref{Eq:LinHam},\ref{Eq:LinMom},\ref{Eq:LinC},\ref{Eq:WeylE},\ref{Eq:WeylB}), respectively. By the assumption, they lie in $C^\infty_0(\Sigma)$. These quantities can be extended to $M$ by solving the well-posed IBVP described in Lemma \ref{Lem:GaugeInv}.

In a next step, we define the quantities $a := -f_0^{-1}\grad\alpha$, $C:=B + G_0^{-1}\kappa_0 a/(f_0-\kappa_0)$, $F:=G_0^{-1}[\divr\beta+ \kappa_0 K/(f_0 -\kappa_0)]$ and $R := G_0^{-1}\curl\beta$ at $t=0$, and extend the fields $\alpha$, $\beta$, $K$, $a$, $C$, $R$, $F$ to $M$ by solving the well-posed IBVP described in Lemma \ref{Lem:Gauge}. Define $B(t):=C(t) - G_0^{-1}\kappa_0 a(t)/(f_0-\kappa_0)$ for all $t > 0$. According to the statement of this lemma, the above definitions for $a$, $C$, $F$ and $R$ are preserved, i.e. they are valid on $M$. 

Next, define for $t > 0$,
\begin{eqnarray}
\Gamma(t) &:=& \Gamma(0) + \frac{B(t) - B(0)}{H_0},\\
A(t) &:=& A(0) + \int\limits_0^t \left[ {\cal E}(s) - \grad\grad\alpha(s) \right] ds,\\
\phi(t) &:=& \phi(0) - \frac{1}{6}\int\limits_0^t \left[ K(s) - \divr\beta(s) \right] ds,\\
\gamma(t) &:=& \gamma(0) - 2\int\limits_0^t \left[ A(s) - \grad\beta(s) \right] ds.
\end{eqnarray}
We claim that the smooth fields $\alpha(t)$, $K(t)$, $\beta(t)$, $B(t)$, $\phi(t)$,
$\gamma(t)$, $A(t)$ 
and $\Gamma(t)$ obtained in this way solve the linearized BSSN equations
(\ref{Eq:LinBSSN1}--\ref{Eq:LinBSSN8}). For this, we first claim that $\frac{1}{2}\divr\Gamma - 4\Delta\phi = {\cal H}$, $\divr A - \frac{2}{3}\grad K = {\cal M}$, $\Gamma - \divr\gamma = {\cal C}$, $-\frac{1}{2}\Delta\gamma + \grad\Gamma - 2\grad\grad\phi = {\cal E}$ and that $\curl A = {\cal B}$ for all $t > 0$. In order to prove this, we first notice that these equalities are true for $t=0$ by definition of the gauge-invariant quantities ${\cal H}$,
${\cal M}$, ${\cal C}$, ${\cal E}$, ${\cal B}$ at $t=0$. Then, the Eqs.~(\ref{Eq:GaugeInv1},\ref{Eq:GaugeInv2},\ref{Eq:GaugeInv3},\ref{Eq:GaugeInv5},\ref{Eq:GaugeInv4bis},\ref{Eq:Gauge3},\ref{Eq:Gauge5}) and the above definitions for $\Gamma$, $A$, $\phi$ and $\gamma$ imply that
\begin{eqnarray}
&& \frac{d}{dt}\left( {\cal H} - \frac{1}{2}\divr\Gamma + 4\Delta\phi \right) = 0,\\
&& \frac{d}{dt}\left( {\cal M} - \divr A + \frac{2}{3}\grad K \right) = 0,\\
&& \frac{d}{dt}\left( {\cal C} - \Gamma + \divr\gamma \right)
 = 2\left( {\cal M} - \divr A + \frac{2}{3}\grad K \right) ,\\
&& \frac{d}{dt} \left( {\cal E} + \frac{1}{2}\Delta\gamma - \grad\Gamma 
+ 2\grad\grad\phi \right)
\nonumber\\
 && \qquad = -\curl\left( {\cal B} - \curl A \right) 
+ \frac{3}{2}\grad\left( {\cal M} - \divr A + \frac{2}{3}\grad K \right)
 + n\otimes\left[ n \wedge \left( \divr{\cal B} - \frac{1}{2}\curl{\cal M} \right) \right],\\
&& \frac{d}{dt}\left( {\cal B} - \curl A \right) = 0,\\
&& \frac{d}{dt} \left(  \divr{\cal B} - \frac{1}{2}\curl{\cal M} \right) = 0,
\end{eqnarray}
which proves the first claim. Now it is straightforward to verify that the fields $\alpha(t)$,
$K(t)$, $\beta(t)$, $B(t)$, $\phi(t)$, $\gamma(t)$, $A(t)$ and $\Gamma(t)$ satisfy the linearized BSSN equations (\ref{Eq:LinBSSN1}--\ref{Eq:LinBSSN8}).

{\em Uniqueness}: Suppose $\alpha^{(1)}$, $K^{(1)}$, $\beta^{(1)}$, $B^{(1)}$, $\phi^{(1)}$, $\gamma^{(1)}$, $A^{(1)}$, $\Gamma^{(1)}$ and $\alpha^{(2)}$, $K^{(2)}$, $\beta^{(2)}$, $B^{(2)}$, $\phi^{(2)}$, $\gamma^{(2)}$, $A^{(2)}$, $\Gamma^{(2)}$ are two smooth solutions of the linearized BSSN equations (\ref{Eq:LinBSSN1}--\ref{Eq:LinBSSN8}) which are identical on the initial surface $t=0$ and which both satisfy the boundary conditions (\ref{Eq:LinMomBC}--\ref{Eq:LinBetaOBC}) with identical data $G_{\perp\perp}$ and $g_\perp$. Then, $\alpha:=\alpha^{(2)}-\alpha^{(1)}$, $\beta:=\beta^{(2)}-\beta^{(1)}$, ... , $\Gamma:=\Gamma^{(2)}-\Gamma^{(1)}$ is also a solution with trivial initial data satisfying the boundary conditions (\ref{Eq:LinMomBC}--\ref{Eq:LinBetaOBC}) with homogeneous boundary data. In particular, the associated gauge-invariant quantities ${\cal H}$, ${\cal M}$, ${\cal C}$, ${\cal E}$, ${\cal B}$ defined by Eqs. (\ref{Eq:LinHam},\ref{Eq:LinMom},\ref{Eq:LinC},\ref{Eq:WeylE},\ref{Eq:WeylB}) satisfy the IBVP described in Lemma \ref{Lem:GaugeInv} with trivial initial and boundary data. Since this IBVP is well posed, it follows that these quantities are zero on $M$. Similarly, it follows from the well-posedness of the IBVP described in Lemma \ref{Lem:Gauge} that $\alpha$, $\beta$, $K$ and $B$ must be zero on $M$. Next, it follows from Eqs. (\ref{Eq:LinBSSN5},\ref{Eq:LinBSSN8}) that $\phi$ and $\Gamma$ are zero on $M$. Next, since $0 = -2{\cal E} = \Delta\gamma$, it follows from Eq.~(\ref{Eq:LinBSSN7}) that $A=0$ on $M$. Finally, Eq.~(\ref{Eq:LinBSSN6}) implies that
also $\gamma=0$ on $M$ which shows that the two solutions are identical.

{\em Constraint preservation}: This will be proven in the next section for the more general
case of the full nonlinear BSSN equations.
\qed

\subsection{Geometric uniqueness}
\label{SubSec:Uniqueness}

In this subsection we analyze the residual gauge freedom in our initial-boundary value
formulation of the linearized BSSN system. With respect to an infinitesimal coordinate transformation generated by a vector field $X^\mu$, say, the linearized four metric $h_{\mu\nu}$ transforms according to
\begin{displaymath}
h_{\mu\nu} \mapsto h'_{\mu\nu} = h_{\mu\nu} + \pounds_X \eta_{\mu\nu} 
 = h_{\mu\nu} + \partial_\mu X_\nu + \partial_\nu X_\mu,
\end{displaymath}
where $\eta_{\mu\nu}$ is the Minkowski metric. This induces the following transformations on the linearized BSSN variables,
\begin{eqnarray}
&& \alpha' = \alpha + F,\qquad
 K' = K - \Delta E, \qquad
 \beta' = \beta + Y,\qquad
 B' = B + Z,
\label{Eq:GaugeTransf1}\\
&& \phi' = \phi + \frac{1}{6}\divr X,\qquad
 \gamma' = \gamma + 2\grad X,\qquad
 A' = A - \grad\grad E,
\label{Eq:GaugeTransf2}\\
&& \Gamma' = \Gamma + \Delta X + \frac{1}{3}\grad\divr X,
\label{Eq:GaugeTransf3}
\end{eqnarray}
where we have defined $E:=X^0$, $X := (X^1,X^2,X^3)$ and $F:=\dot{E}$, $Y:=\dot{X} - \grad E$, $Z:=G_0^{-1}\dot{Y}$. The linearized BSSN equations (\ref{Eq:LinBSSN1}--\ref{Eq:LinBSSN8}) are invariant with respect to these transformations if and only if $E$, $F$, $Y$ and $Z$ satisfy the equations
\begin{eqnarray}
\dot{E} &=& F,
\label{Eq:GaugeEvolution1}\\
\dot{F} &=& f_0\Delta E,
\label{Eq:GaugeEvolution2}\\
\dot{Y} &=& G_0 Z,
\label{Eq:GaugeEvolution3}\\
\dot{Z} &=& H_0\left( \Delta Y + \frac{1}{3}\grad\divr Y + \frac{4}{3f_0}\grad\dot{F} \right).
\label{Eq:GaugeEvolution4}
\end{eqnarray}
Similarly, the boundary conditions (\ref{Eq:LinMomBC}--\ref{Eq:LinBetaOBC}) are invariant 
provided that $F$ and $Y$ satisfy
\begin{equation}
\partial_{||} F \hateq 0,\qquad
Y_{||} \hateq 0,\qquad
\left( \partial_t + \frac{\sqrt{3\kappa_0}}{2}\,\partial_{||} \right) Y_{\perp} 
 \hateq \frac{\kappa_0}{f_0 - \kappa_0}\,\partial_{\perp} F + G_0 w_\perp,
\label{Eq:GaugeBC}
\end{equation}
where $w_\perp = g'_\perp - g_\perp$ and $g_\perp$ and $g'_\perp$ are the boundary data corresponding to  the two different gauges. Since the Eqs.~(\ref{Eq:GaugeEvolution1}--\ref{Eq:GaugeEvolution4}) yield the two second-order equations
\begin{equation}
\ddot{F} = f_0\Delta F, \qquad
\ddot{Y} = G_0 H_0\left( \Delta Y + \frac{1}{3}\grad\divr Y + \frac{4}{3f_0}\grad\dot{F} \right)
\label{Eq:GaugeEvolution2nd}
\end{equation}
which are identical to the second-order equations for $\alpha$ and $\beta$ obtained from 
Eqs.~(\ref{Eq:LinBSSN1}--\ref{Eq:LinBSSN4}), and since the boundary conditions
(\ref{Eq:GaugeBC}) are identical to the boundary conditions
(\ref{Eq:LinLapseBC}--\ref{Eq:LinBetaOBC}), it follows from the results in Lemma
\ref{Lem:Gauge} that the IBVP for $F$ and $Y$ consisting of the evolution equations
(\ref{Eq:GaugeEvolution2nd}), the boundary conditions (\ref{Eq:GaugeBC}), initial data for $F$, $\dot{F}=f_0\Delta E$, $Y$, $\dot{Y} = G_0 Z$ in $L^2(\Sigma)$ and boundary data $w_\perp$ in $L^2({\cal T})$, is well posed. Therefore, the residual gauge freedom is entirely determined by the initial values for $E$, $F$, $Y$, $Z$ and the boundary data $g_\perp$ and $g'_\perp$. This allows us to formulate the following result:

\begin{theorem}[Geometric Uniqueness]
\label{Thm:Uniqueness}
Two smooth solutions of the IBVP described in Theorem \ref{Thm:Main} are related to each other by an infinitesimal coordinate transformation if and only if there exists two smooth functions $F$ and $E$ on $\Sigma$, three smooth vector fields $X$, $Y$ and $Z$ on $\Sigma$ and a smooth vector field $w_\perp$ on ${\cal T}$ such that the initial data of the two solutions are related to each other by the transformations
(\ref{Eq:GaugeTransf1},\ref{Eq:GaugeTransf2},\ref{Eq:GaugeTransf3}) and the boundary data $(G_{\perp\perp},g_\perp)$, $(G'_{\perp\perp},g'_\perp)$ is related to each other by
\begin{displaymath}
G'_{\perp\perp} \hateq G_{\perp\perp},\qquad
g'_\perp \hateq g_\perp + w_\perp.
\end{displaymath}
\end{theorem}

\proof The "only if" part of the proof follows from the considerations above. For the "if"
part, we use the fields $E$, $F$, $Y$ and $Z$ to set up initial data for $F$, $\dot{F}=f_0\Delta E$, $Y$, $\dot{Y} = G_0 Z$, and the field $w_\perp$ to set up boundary data for the IBVP described by Eqs.~(\ref{Eq:GaugeEvolution2nd}) and (\ref{Eq:GaugeBC}). Since this problem is well posed, we may extend the fields $F$ and $Y$ uniquely to all $M$. Defining
\begin{displaymath}
E(t) := E(0) + \int\limits_0^t F(s) ds, \qquad
X(t) := X(0) + \int\limits_0^t \left[ Y(s) + \grad E(s) \right] ds,
\end{displaymath}
we obtain a vector field $(X^\mu(t)) = (E(t),X(t))$ on $M$ which, by construction, generates an infinitesimal coordinate transformation between two solutions of the linearized IBVP for the BSSN system.
\qed

\section{Constraint preservation}
\label{Sec:CP}

In this section we prove that smooth enough solutions satisfying the nonlinear BSSN
equations (\ref{Eq:BSSN1}--\ref{Eq:BSSN8}) and our boundary conditions satisfy the nonlinear constraints (\ref{Eq:BSSNCons1},\ref{Eq:BSSNCons2},\ref{Eq:BSSNCons3})
everywhere on the spacetime domain $M = [0,T]\times\Sigma$ if the initial data satisfy these constraints on the  initial slice $\{ 0 \} \times \Sigma$. The proof is based on the FOSH system derived in \cite{dBoSeSmTpDiHdP07,dBpDoSeSmT09} governing the propagation of the constraint fields with the parameter choice $m=1$, $\sigma=1/2$. The detailed equations describing this system will not be needed for the following argument. Only its form,
\begin{equation}
\hat{\partial}_0 C = \alpha
\left[ {\bf A}(u)^i\partial_i C + {\bf B}(u) C \right],
\label{Eq:ConstrEvol}
\end{equation}
is important. Here, $C = ({\cal C}^i,{\cal H}, S := {\cal H} + Z/2, {\cal M}_j, \hat{Z}_{(ij)},
Z_{[ij]})$ are the constraint variables, where we have decomposed the variable $Z_{ij} := (\partial_i {\cal C}^k)\tilde{\gamma}_{kj} = \hat{Z}_{(ij)} + Z_{[ij]} + \gamma_{ij} Z/3$ into its trace-free symmetric part, $\hat{Z}_{(ij)}$, its antisymmetric part, $Z_{[ij]}$, and its trace, $Z = \gamma^{ij} Z_{ij}$, where $u = (\alpha,\beta^i,\phi,K,\tilde{\gamma}_{ij},\tilde{A}_{ij})$ are the main variables, and where ${\bf A}^i$, $i=1,2,3$, and ${\bf B}$ are matrix-valued functions of $u$. The principal symbol ${\bf A}({\bf n}) = {\bf A}(u)^i n_i$ is given by
\begin{equation}
{\bf A}({\bf n})\left( \begin{array}{c} 
 {\cal C}^i \\ {\cal H} \\ S \\ {\cal M}_j \\ \hat{Z}_{(ij)} \\ Z_{[ij]})
\end{array} \right)
 = \left( \begin{array}{c}
 0 \\ 0 \\ n^j {\cal M}_j \\
 \frac{1}{3} n_j S + \frac{1}{2} n^i \hat{Z}_{(ij)} + \frac{1}{2} n^i Z_{[ij]} \\
 2  (n_{(i} {\cal M}_{j)})^{TF} \\
 2\, n_{[i} {\cal M}_{j]}
\end{array} \right) .
\label{Eq:CPSymbol}
\end{equation}
Here $n^i \equiv \gamma^{ij} n_j$ and $n_i$ is normalized such that $n_i n^i = 1$. This system is symmetric hyperbolic;  a symmetrizer ${\bf H} = {\bf H}^T$ is given by the quadratic form
\begin{displaymath}
C^T {\bf H} C = \tilde{\gamma}_{ij} {\cal C}^i {\cal C}^j + {\cal H}^2
 + \frac{1}{3}\, S^2 + \gamma^{ij} {\cal M}_i {\cal M}_j
 + \frac{1}{4}\,\gamma^{ik}\gamma^{jl}\hat{Z}_{(ij)}\hat{Z}_{(kl)}
 + \frac{1}{4}\,\gamma^{ik}\gamma^{jl} Z_{[ij]} Z_{[kl]}\; .
\nonumber
\end{displaymath}
In order to obtain the class of maximal dissipative boundary conditions, we find, following
the appendix, the characteristic speeds and fields which are defined as the eigenvalues and corresponding projections of $C$ onto the eigenspaces of ${\bf A}(n)$. They are given by
\begin{eqnarray}
& \mu = 0,
& {\cal C}^k\, , \qquad 
  {\cal H}\, , \qquad
 \frac{4}{3} S - n^i n^j\hat{Z}_{(ij)}\, ,\qquad
 (\hat{Z}_{(kj)} + Z_{[kj]})\Pi^k_i n^j\, ,\qquad
 Z_{[kl]} \Pi^k_i\Pi^l_j\, ,\qquad
 \hat{Z}_{(kl)}P^{kl}{}_{ij}\, ,
\nonumber\\
& \mu = \pm 1, 
& V^{(\pm 1)} = n^j {\cal M}_j \pm
\left[ \frac{1}{3}\, S + \frac{1}{2} n^i n^j\hat{Z}_{(ij)} \right],
\nonumber\\
& \mu = \pm 1,
& W^{(\pm 1)}_j = {\cal M}_k\Pi^k_j  
\pm n^i\left[ \frac{1}{2} \hat{Z}_{(ik)} + \frac{1}{2} Z_{[ik]} \right]\Pi^k_j ,
\nonumber
\end{eqnarray}
where $\Pi^j_k$ and $P^{ij}{}_{kl}$ are the projection operators defined in Table~\ref{Tablita}. In terms of these fields we find
\begin{displaymath}
C^T {\bf H}{\bf A}(n) C = 
\frac{1}{2}\left[ (V^{(+ 1)})^2 - (V^{(- 1)})^2 \right]
+ \frac{1}{2}\gamma^{ij} \left[ W^{(+ 1)}_i W^{(+ 1)}_j - W^{(-1)}_i W^{(-1)}_j \right],
\end{displaymath}
which shows that the boundary conditions
\begin{equation}
V^{(+1)} \hateq b_1 V^{(-1)}, \qquad
W^{(+1)}_j \hateq b_2 W^{(-1)}_j,
\end{equation}
where $b_1$ and $b_2$ are two functions on the boundary ${\cal T}$ whose magnitude is smaller than or equal to one, and where $n_i$ refers to the outward unit normal to $\partial\Sigma$, are in maximal dissipative form. The particular case $b_1 = b_2 = -1$ corresponds to imposing the momentum constraint on the boundary, ${\cal M}_j \hateq 0$. Since the constraint evolution system (\ref{Eq:ConstrEvol}) is linear and homogeneous, this implies that the unique solution with trivial initial data is zero. Therefore, imposing the momentum constraint on the boundary guarantees that solutions which satisfy the constraints initially automatically satisfy them on $M$. Furthermore, the energy estimate from the appendix shows that a suitable norm of the constraint violations which are usually present in numerical calculations is bounded by a (time-dependent) constant times the norm of the initial constraint violations. Therefore, for each fixed time $t$, smaller and smaller initial constraint violations lead to smaller and smaller constraint violations at $t$.
Finally, we mention that our boundary conditions do not modify the key property of the turducken approach \cite{dBoSeSmTpDiHdP07,dBpDoSeSmT09} regarding the causal propagation of constraint violations. In particular, constraint violations initiating inside a black hole cannot propagate outside the hole for the parameter choice $m=1$ used in the present work.

\section{A note on the numerical implementation}
\label{Sec:Numer}

Regarding the numerical implementation of our boundary conditions we have the following
comments: from the system of evolution equations, Eqs.~(\ref{Eq:BSSN1}--\ref{Eq:BSSN8}), it is clear that a question arises at the moment of discretizing the spatial derivatives at the boundary points. For finite difference methods, one possibility is to use one-sided differences at those points. However, this simple method cannot yield a convergent numerical schemes in general since it does not incorporate the information about the boundary conditions, and so the corresponding system at the continuum is underdetermined. (Using one-sided differences would be enough if the boundary is purely outflow, as happens at the inner boundary of some excision schemes.) The question then is how to incorporate the boundary conditions summarized in Table~\ref{Tablita} into the finite difference discretization. Without specifying a detailed method for doing so, let us nevertheless mention a few ideas.

Concerning the gauge boundary conditions $\partial_n\alpha = 0$, $\beta^n=0$ on the lapse and the normal component of the shift, they could be used in order to extrapolate the values of $\alpha$ and $\beta^n$ to ghost zones outside the numerical grid. These values could then be used in order to compute centered differences at the boundary points. A similar procedure could be applied to the gauge condition for the tangential components of the shift.

In order to implement the constraint preserving and $\Psi_0$ specifying conditions, first note that they are somewhat less common than the gauge boundary ones, as they involve second-order derivatives of the metric fields. In this case, we propose to follow a procedure described in \cite{oSmT05}, where constraint preserving boundary conditions, including the $\Psi_0$ specifying condition, are numerically implemented and tested for a first-order symmetric hyperbolic formulation of Einstein's equations. Applied to the BSSN system, this method consists in adding terms to the right hand side of the evolution equations, Eqs.~(\ref{Eq:BSSN1}--\ref{Eq:BSSN8}), which are proportional to the 
expressions $\Phi_i \equiv \tilde{D}^j \tilde{A}_{ij} - \frac{2}{3} \tilde{D}_i K +
6\tilde{A}_{ij} \tilde{D}^j\phi - 8\pi G_N j_i$ and $\Psi_{lm} \equiv {P^{ij}}_{lm}{\bar{\cal E}}_{ij} - \left({n^i\,P^{kj}}_{lm} - n^k\,{P^{ij}}_{lm}\right)\,{\bar{\cal B}}_{kij} -
{P^{ij}}_{lm}\,G_{ij}$, defining the momentum constraint and the $\Psi_0$ specifying boundary conditions, respectively. The proportionality factors are then chosen in such a way that normal derivatives corresponding to the combinations of the characteristic fields appearing in the expressions for $\Phi_i$ and $\Psi_{ij}$, are eliminated. The characteristic fields for the BSSN Eqs.~(\ref{Eq:BSSN1}--\ref{Eq:BSSN8}) have been calculated in \cite{hBoS04} based on the pseudo-differential approach described in \cite{KreissOrtiz-2002,gNoOoR04}. The ones that are relevant to our problem are
\begin{eqnarray}
v^{(\pm 1)} &=& e^{4\phi} n^i n^j\tilde{A}_{ij} - \frac{2K}{3} 
 \mp n^k\left( \frac{1}{2} e^{4\phi} n^i n^j\partial_k\tilde{\gamma}_{ij} -
\frac{2}{3}\tilde{\Gamma}_k 
+ \frac{4}{3}\partial_k\phi \right),
\nonumber\\
v^{(\pm 1)}_k &=& e^{4\phi} n^i\Pi^j_k\tilde{A}_{ij} \mp \Pi^l_k\left( \frac{1}{2} e^{4\phi} 
n^i n^j\partial_i\tilde{\gamma}_{jl} - \frac{1}{2}\tilde{\Gamma}_l \right),
\nonumber\\
v^{(\pm 1)}_{kl} &=& e^{4\phi} {P^{ij}}_{kl}\left( \tilde{A}_{ij} 
\mp \frac{1}{2} n^s\partial_s\tilde{\gamma}_{ij} \right),
\nonumber
\end{eqnarray}
where $n$ refers to the unit outward normal to the boundary, and the projectors $\Pi^i_j$ 
and ${P^{ij}}_{kl}$ are defined in Table~\ref{Tablita}. In terms of these fields, we have
\begin{eqnarray}
\Phi_j &=& \frac{1}{2}\partial_n\left( v^{(+1)} + v^{(-1)} \right) n_j 
 + \frac{1}{2}\partial_n\left( v_j^{(+1)} + v_j^{(-1)} \right) + ...,\\
\Psi_{ij} &=& \partial_n v_{ij}^{(+1)}  + ...,
\end{eqnarray}
where $...$ refers to terms containing tangential derivatives or undifferentiated combinations of the characteristic fields only. Therefore, at the boundary points, $\Phi_j$ and $\Psi_{ij}$ have to be added to the evolution equations in such a way that the normal
derivatives of $v^{(+1)} + v^{(-1)}$, $v_j^{(+1)} + v_j^{(-1)}$ and $v_{ij}^{(+1)}$ are eliminated.

Once this calculation has been performed and the coefficients in front of  $\Phi_i$ and $\Psi_{ij}$ have been determined, the BSSN evolution equations 
(\ref{Eq:BSSN1}--\ref{Eq:BSSN8}) with the additional boundary terms may be discretized using finite difference operators. At the boundary, one-sided differences might be used for all fields except for lapse and shift for which centered differences with the extrapolated values at the ghost zones are implemented, as described above. We stress that once the above mentioned coefficients have been determined, the characteristic fields are not required anymore and the system might be implemented numerically.

For a different method for discretizing constraint-preserving boundary conditions based on 
pseudo-spectral methods we refer the reader to \cite{lKlLmSlBhP05}.

\section{Conclusions}
\label{Sec:Conc}

In this work we have analyzed the IBVP for the BSSN evolution system of Einstein's field
equations with a hyperbolic $K$-driver and Gamma-driver condition \cite{mAetal03} for lapse and shift, as used in current numerical simulations. Unlike the harmonic formulation which has been motivated by the mathematical structure of the equations and the understanding of the Cauchy formulation in General Relativity, the BSSN system has been developed and improved based on its capability of numerically evolving binary black hole spacetimes in a stable way. Therefore, it is not evident at all that mathematical questions like the well-posedness of the IBVP may be answered.

One of the important results obtained in this article is that the BSSN evolution system yields a "nice" evolution system for the constraint variables and the Weyl curvature and a "nice" evolution system for lapse and shift. These two systems are "nice" in the sense that they may be cast into first-order symmetric hyperbolic form, from which one concludes that they yield a well-posed time evolution with standard prescriptions for constructing boundary conditions. At the linearized level, the two system decouple from each other and the gauge-dependent fields may be constructed from their solutions as we have shown in Theorem~\ref{Thm:Main}. At the nonlinear level, these systems couple to each other through the metric fields, and one needs to consider the coupled system together with the evolution of the metric fields. If this large system could be cast into first-order symmetric hyperbolic form as well, a well-posedness proof would follow from the standard theorems quoted in the appendix. We have not investigated this issue in the present article. Instead, we have extrapolated the boundary conditions constructed in the linearized case to the full nonlinear BSSN system. These conditions are summarized in Table~\ref{Tablita}, and they are conjectured to yield a well-posed system in the nonlinear case as well.

Summarizing, our new boundary conditions for the nonlinear BSSN system have the following properties: they preserve the constraints throughout evolution, as shown in Sec.~\ref{Sec:CP}, they yield a well-posed IBVP in the linearized case, and furthermore, they control the Weyl scalar $\Psi_0$ at the boundary, a condition that is currently used in binary black hole simulations based on the harmonic formulation of Einstein's equations (see, for instance, Ref.~\cite{hPdBlKlLgLmS07}) and has been tested for its accuracy for nonlinear waves on a Schwarzschild background \cite{oRlLmS07}.

In a first step towards their actual numerical implementation, we have sketched a possible 
numerical method for imposing our boundary conditions which is based on finite-differences and projection techniques. It is the hope of the authors that such an implementation eventually improves the accuracy of the simulations and avoids the need of pushing the boundary so far away that it is causally disconnected from the region where physics is extracted. Our boundary conditions may also be useful for matching the solution at interface boundaries, or for the Cauchy-characteristic matching approach.


\acknowledgments

We wish to thank H. Beyer, J. Gonz\'alez, F. Guzm\'an, S. Husa, C. Palenzuela, O. Reula, J. Winicour, T. Zannias  and A. Zengino\u{g}lu for enlightening comments and discussions. DN acknowledges partial support from the DAAD, and DGAPA-UNAM and from CONACyT Grant No. U-47209-F. OS was supported in part by Grants No. CIC 4.19 to Universidad Michoacana, PROMEP UMICH-PTC-195 from SEP Mexico, and CONACyT 61173.

\appendix
\section{Symmetric hyperbolic first-order systems with boundaries}
\label{App}

In this appendix we summarize some known results about FOSH systems with maximal 
dissipative boundary conditions \cite{kF58,pLrP60}. Let $\Sigma\subset \Real^n$ be an open subset of $\Real^n$ with closure $\bar{\Sigma}$ and $C^\infty$ boundary $\partial\Sigma$. We consider the following IBVP on the spacetime region $M:=[0,\infty) \times
\bar{\Sigma}$,
\begin{eqnarray}
& \partial_t u = A^i(t,x)\partial_i u + D(t,x)u + F(t,x), & t > 0, x\in\Sigma,
\label{Eq:SymHypEq}\\
& u(0,x) = u_0(x), & x\in\Sigma,
\label{Eq:SymHypID}\\
& B(t,x) u(t,x) = G(t,x), & t > 0, x\in\partial\Sigma.
\label{Eq:SymHypBC}
\end{eqnarray}
Here, $A^1$, $A^2$, ... ,$A^n$, $D: M \to Mat(m\times m,\Real)$ are $C^\infty$ matrix-valued  functions on $M$, $F: M \to \Real^m$ is a $C^\infty$ vector-valued function on $M$ and $B: {\cal T} \to Mat(m\times m,\Real)$ is a $C^\infty$ matrix-valued function on the boundary ${\cal T} := [0,\infty) \times \partial\Sigma$. The data consists of the initial data $u_0\in L^2(\Sigma)$ and the boundary data $G\in L^2({\cal T})$, and $u: M \to \Real^m$ is the solution vector, lying in an appropriate function space.

We require the following conditions on the principal symbol $A(t,x,k) := A^i(t,x) k_i$, 
$k = (k_1,k_2,...,k_n)\in \Real^n$, and the boundary matrix $B(t,x)$:
\begin{enumerate}
\item[(a)] There exists a $C^\infty$ matrix-valued function $H: M\to Mat(m\times m,\Real)$, 
called the symmetrizer, such that
\begin{enumerate}
\item[(i)] $H(t,x) = H(t,x)^T$ is symmetric for all $(t,x)\in M$.
\item[(ii)] There exists a constant $C > 0$ such that $C^{-1} |u|^2 \leq u^T H(t,x) u \leq
C|u|^2$ for all $(t,x)\in M$ and all $u\in \Real^m$, that is, $H$ is uniformly positive definite.
\item[(iii)] $H(t,x) A(t,x,k) = A(t,x,k)^T H(t,x)$ is symmetric for all $(t,x)\in M$ and all $k\in \Real^n$.
\end{enumerate}
\item[(b)] Let $J(t,x):=H(t,x) A(t,x,k)$, where $(t,x)\in {\cal T}$ and where $k$ denotes 
the outward unit normal to $\partial\Sigma$. Then, we assume that for each $p = (t,x)\in {\cal
T}$ the boundary space
\begin{displaymath}
V_p := \{ u\in \Real^m : B(t,x) u = 0 \} \subset \Real^m
\end{displaymath}
is maximal non-positive with respect to $J(t,x)$, that is,
\begin{enumerate}
\item[(iv)] $u^T J(t,x) u \leq 0$ for all $u\in V_p$.
\item[(v)] $V_p$ is maximal with respect to the condition (iv), that is, if $W \supset V_p$ is 
a linear subspace of $\Real^m$ containing $V_p$ which satisfies (iv), then $W=V_p$.
\end{enumerate}
Furthermore, we require
\begin{enumerate}
\item[(vi)] The rank of $J(t,x)$ is constant.
\end{enumerate}
\end{enumerate}
Under these conditions, it can be shown that the IBVP (\ref{Eq:SymHypEq},\ref{Eq:SymHypID},\ref{Eq:SymHypBC}) is well posed \cite{pLrP60,jR85,pS96a}. In \cite{pLrP60} a generalization to other domains $\Sigma$ with boundaries which might contain corners is also given. Generalizations to quasi-linear systems, where $A^i(t,x,u)$, $D(t,x,u)$ and $B(t,x,u)$ might also depend on the solution vector itself are also possible \cite{pS96b}.

The meaning of the conditions (i-iv) is that they yield an a priori energy estimate. In order 
to see this, let $u$ be a solution of the IBVP
(\ref{Eq:SymHypEq},\ref{Eq:SymHypID},\ref{Eq:SymHypBC}). By redefining $u$ if necessary, we might assume that $G=0$. Next, define the energy norm
\begin{displaymath}
E(t) := \int\limits_\Sigma u(t,x)^T H(t,x) u(t,x) d^n x 
 = \| H^{1/2}(t,\cdot) u(t,\cdot) \|_{L^2(\Sigma)}^2, \qquad t\geq 0.
\end{displaymath}
Taking a time derivative and using the symmetry of $H$ and the evolution equations (\ref{Eq:SymHypEq}) we obtain
\begin{displaymath}
\frac{d}{dt} E(t) = \int\limits_\Sigma\Big\{
 2 u(t,x)^T H(t,x)\left[ A^i(t,x)\partial_i u + D(t,x)u + F(t,x) \right]
 + u(t,x)^T[\partial_t H(t,x) ] u(t,x) \Big\} d^n x.
\end{displaymath}
Next, using the symmetry of $H(t,x) A^i(t,x)$ and Gauss's theorem, this yields
\begin{displaymath}
\frac{d}{dt} E(t) = \int\limits_{\partial\Sigma} u(t,x)^T J(t,x) u(t,x) d\sigma
 + \int\limits_{\Sigma} u(t,x)^T L(t,x) u(t,x) d^n x
 +  2\int\limits_{\Sigma} u(t,x)^T H(t,x) F(t,x) d^n x,
\end{displaymath}
where $d\sigma$ denotes the area element on $\partial\Sigma$ and where $L(t,x) := \partial_t H(t,x) - \partial_i[H(t,x) A^i(t,x)] + H(t,x) D(t,x) + D(t,x)^T H(t,x)$. The first term on the right-hand side is non-positive due to condition (iv). Using Schwarz' inequality and the
condition (ii) the second and third term on the right-hand side can be estimated as
\begin{displaymath}
\int\limits_{\Sigma} u(t,x)^T L(t,x) u(t,x) d^n x \leq a(t) \| u(t,.) \|_{L^2(\Sigma)}^2
\leq C a(t) E(t),
\end{displaymath}
where $a(t):=\sup\{ \| L(s,x) \| : 0\leq s \leq t, x\in\Sigma \}$, and
\begin{displaymath}
2\int\limits_{\Sigma} u(t,x)^T H(t,x) F(t,x) d^n x
  \leq 2 E(t) \sqrt{C} \| F(t,.) \|_{L^2(\Sigma)} 
  \leq  \frac{C}{K^2} E(t) + K^2\| F(t,.) \|_{L^2(\Sigma)}^2,
\end{displaymath}
where $K > 0$ is a constant, respectively. Summarizing, we obtain
\begin{displaymath}
\frac{d}{dt} E(t) \leq b(t) E(t) + K^2\| F(t,.) \|_{L^2(\Sigma)}^2,
\end{displaymath}
where $b(t) := C[a(t) + K^{-2}]$. Finally, using Gronwall's lemma we have
\begin{equation}
E(t) \leq e^{b(t)t} E(0) + K^2\int\limits_0^t e^{b(t)(t-s)}\| F(s,.) \|_{L^2(\Sigma)}^2 ds,
\qquad t\geq 0.
\end{equation}
Such or similar a priori estimates are the basis for showing different properties of the 
solutions, like their uniqueness, continuous dependence on the data and the finite speed of propagation. The condition (v) is important for the existence of solutions, ensuring that not "too many" boundary conditions are specified.

A more practical formulation of the conditions (iv) and (v) is the following: 
let $p = (t,x)\in {\cal T}$ be fixed, and let $k$ denote the outward unit normal to
$\partial\Sigma$ at $x$. Since $A(t,x,k)$ is symmetric with respect to the scalar product on
$\Real^m$ defined by $< u,v> := u^T H(t,x) v$, $u,v\in\Real^m$, there exists a basis
$e_1,e_2,...,e_m$ of eigenvectors of $A(t,x,k)$ which is orthonormal with respect to
$<\cdot,\cdot>$. Let $\lambda_1,\lambda_2,...,\lambda_m$ be the corresponding eigenvalues, where we might assume that the first $r$ of these eigenvalues are strictly negative, and the last $s$ are strictly positive. We can expand any vector $u\in\Real^m$ as
\begin{displaymath}
u = \sum\limits_{j=1}^m u_j e_j.
\end{displaymath}
The coefficients $u_j$ are called the characteristic fields; the associated speeds are given by $\lambda_j$. Then, we have
\begin{displaymath}
u^T J(t,x) u = < u, A(t,x,k) u > 
 = \sum\limits_{j=1}^m \lambda_j u_j^2
 = -\sum\limits_{j=1}^r |\lambda_j| u_j^2 + \sum\limits_{j=m-s+1}^m \lambda_j u_j^2.
\end{displaymath}
In our work, the nonzero eigenvalues come in pairs, such that $r=s$ and $\lambda_1 =
-\lambda_m$, 
$\lambda_2 = -\lambda_{m-1}$, ... ,$\lambda_r = -\lambda_{m-r+1}$. In this case the above simplifies to
\begin{displaymath}
u^T J(t,x) u = \sum\limits_{j=1}^r |\lambda_j|\left( |v_j^{(+)}|^2 - |v_j^{(-)}|^2 \right),
\end{displaymath}
where we have defined $v_j^{(+)} := u_{m-j-1}$ and $v_j^{(-)}:=u_j$. Clearly, the boundary conditions $v_j^{(+)} = c^{(j)} v_j^{(-)}$ with $|c^{(j)}| \leq 1$, $j=1,2,...,r$, satisfy the conditions (iv) and (v).

\bibliographystyle{unsrt}
\bibliography{refs}

\end{document}